%% file: DA_position_space.tex
\documentclass[aps,prd,a4paper,longbibliography,reprint,superscriptaddress,amsmath,amssymb,showpacs]{revtex4-1}
\usepackage{graphicx}
\usepackage{subfigure}

\usepackage{mathtools}
\usepackage{MnSymbol}
\usepackage{dsfont}
\usepackage{bbold}
\usepackage{slashed}

\usepackage{marvosym}

\usepackage{lipsum}
\usepackage{xcolor}
\usepackage{colortbl}
\usepackage{dcolumn}

\usepackage[plainpages=false,pdfpagelabels,pageanchor=false]{hyperref}

\usepackage[squaren]{SIunits}

\usepackage{simplewick}

\newcommand{\overbar}[1]{\mkern 1.5mu\overline{\mkern-1.5mu#1\mkern-1.5mu}\mkern 1.5mu}
\newcommand{\MSbar}{\overbar{\text{MS}}}
\newcommand\mydots{\hbox to 1em{.\hss.\hss.}}

\newcommand{\jhat}{\skew{4}{\hat}{\boldsymbol{\jmath}}}
\DeclareTextSymbolDefault{\textperiodcentered}{OMS}
\renewcommand{\cdot}{\mathbin{\mbox{\textperiodcentered}}}

\newcolumntype{E}[1]{D{.}{.}{#1}}
\makeatletter
\def\DC@endright{$\hfil\egroup\@dcolcolor\box\z@\box\tw@\dcolreset}
\def\dcolcolor#1{\gdef\@dcolcolor{\color{#1}}}
\def\dcolreset{\dcolcolor{black}}
\dcolcolor{black}
\makeatother

\newcommand{\affone}{\affiliation{Institut f{\"u}r Theoretische Physik, Universit{\"a}t Regensburg, Universit{\"a}tsstra{\ss}e 31, 93053 Regensburg, Germany}}
\newcommand{\afftwo}{\affiliation{Department of Theoretical Physics, Tata Institute of Fundamental Research, Homi Bhabha Road, Mumbai 400005, India}}
\newcommand{\affthree}{\affiliation{Zentrum f\"ur Datenverarbeitung, Universit\"at T\"ubingen, W\"achterstra{\ss}e 76, 72074 T\"ubingen, Germany}}
\newcommand{\afffour}{\affiliation{Marian Smoluchowski Institute of Physics, Jagiellonian University, ul.\ {\L}ojasiewicza 11, 30-348 Krak\'ow, Poland}}

\begin{document}
\begin{abstract}%
Building upon our recent study [G.~S.~Bali~\emph{et~al.},~Eur.~Phys.~J.~{\bf{C78}},~217~(2018)], we investigate the feasibility of calculating the pion distribution amplitude from suitably chosen Euclidean correlation functions at large momentum. We demonstrate in this work the advantage of analyzing several correlation functions simultaneously and extracting the pion distribution amplitude from a global fit. This approach also allows us to study higher-twist corrections, which are a major source of systematic error. Our result for the higher-twist parameter~$\delta^\pi_2$ is in good agreement with estimates from QCD sum rules. Another novel element is the use of all-to-all propagators, calculated using stochastic estimators, which enables an additional volume average of the correlation functions, thereby reducing statistical errors.%
\end{abstract}
%\keywords{Lattice QCD calculations,Factorization,Light mesons}

\title{Pion distribution amplitude from Euclidean correlation functions:\texorpdfstring{\\}{ }Exploring universality and higher-twist effects}

\author{Gunnar~S.~\surname{Bali}}\affone\afftwo
\author{Vladimir~M.~\surname{Braun}}\affone
\author{Benjamin~\surname{Gl\"a{\ss}le}}\affone\affthree
\author{Meinulf~\surname{G\"ockeler}}
\author{Michael~\surname{Gruber}}
\author{Fabian~\surname{Hutzler}}\affone
\author{Piotr~\surname{Korcyl}}\affone\afffour
\author{Andreas~\surname{Sch\"afer}}
\author{Philipp~\surname{Wein}}\email[]{philipp.wein@physik.uni-regensburg.de}
\author{Jian-Hui~\surname{Zhang}}\affone
\date{\today}
\maketitle
\section{Introduction}\label{sect_introduction1}%
The lattice approach to QCD enables the computation of a multitude of hadronic parameters with high precision from first principles. Since the inception of this method, the list of quantities amenable to lattice simulation has been ever increasing. As the scientific focus moves on to ever larger classes of quark-gluon correlations the need for high precision lattice simulations to complement experimental data becomes ever more urgent. Hadronic contributions to the muon anomalous magnetic moment, which is on the verge of becoming a sensitive probe of physics beyond the Standard Model, constitute one such prominent example. In particular lattice calculations of the hadronic ``light-by-light'' scattering contribution~\cite{Blum:2017cer,Asmussen:2018ovy} are set to become more precise than inferring this quantity from experimental measurements; see, e.g., Refs.~\cite{Nyffeler:2016gnb,Colangelo:2017fiz} and references therein. Another venue which currently attracts a lot of attention is how lattice QCD may contribute to the determination of parton (i.e., quark and gluon) distributions in hadrons~\cite{Lin:2017snn}, which are scale-dependent nonperturbative quantities that enter the description of ``hard'' processes via QCD factorization theorems.\par
The possibility of calculating parton distributions from Euclidean correlation functions has been discussed for decades. For early work, see, e.g., Refs.~\cite{Liu:1993cv,Aglietti:1998ur,Abada:2001if}. Recently, with the work by Ji~\cite{Ji:2013dva} in which it was strongly emphasized that nothing prevents one from accessing light-cone dynamics starting from Euclidean space, such approaches gained prominence. Several proposals exist that differ in detail but share the same general strategy: the parton distributions are not calculated directly but extracted from suitable Euclidean correlation functions (``lattice cross sections'' in the terminology of Ref.~\cite{Ma:2017pxb}; we prefer to use the term ``Euclidean correlation functions'' in this context because cross sections, in general, do not have a simple path-integral representation). After taking the continuum and other appropriate limits, these can be expressed in terms of parton distributions in the framework of QCD factorization in continuum theory, in analogy to the extraction of parton distributions from fits to experimentally measured structure functions. In other words, the role of lattice QCD can be to provide a complementary set of observables from which parton distributions can be extracted, ideally, employing global fits combining lattice input with experimental data on hard reactions.\par
Such calculations are at an exploratory stage. At present, the main task is to develop specific techniques that will eventually allow one to control all systematic errors. The pion light-cone distribution amplitude (DA) is the simplest parton function of this kind and offers itself as a laboratory where many of the relevant issues can be investigated. It also allows one to compare the strengths and weaknesses of the existing methods. Moreover, the pion DA is interesting in its own right as the main nonperturbative input to studies of hard exclusive reactions with energetic pions in the final state, e.g., the $\gamma^\ast\gamma\to \pi$ transition form factor and weak $B$-meson decays $B\to \pi \ell\nu_\ell$, $B\to \pi\pi$, etc.\par
\begin{figure*}[t]%
\centering
\includegraphics[width=\textwidth,clip]{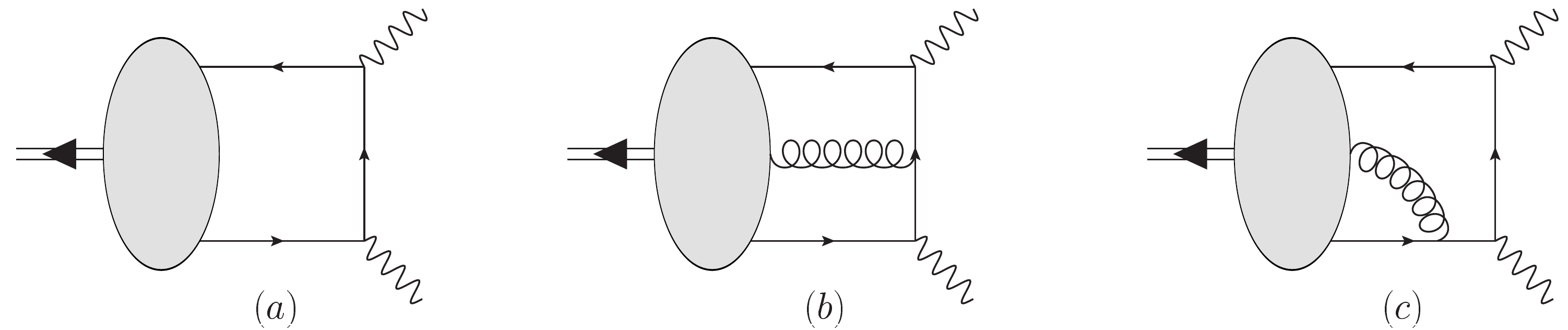}%
\caption{\label{fig-pigammagamma}The leading-twist $(a)$ and higher-twist $(b)$, $(c)$ leading order contributions to the pion transition form factor.}
\end{figure*}%
In our recent publication~\cite{Bali:2017gfr}, we have showcased the position space approach proposed in Ref.~\cite{Braun:2007wv}, and we will be using the same framework here. The new contribution of this work is to illustrate the advantages of considering several correlation functions simultaneously. Such a multichannel approach not only leads to better statistics but also, most importantly, allows one to control and estimate higher-twist corrections which otherwise lead to large systematic errors. The possibility of constraining higher-twist corrections from the studies of lattice correlation functions is interesting within a much more general context and can have important applications. For the case at hand, we find that the higher-twist corrections extracted from lattice simulations agree very well with earlier estimates based on QCD sum rules and the phenomenology of hard exclusive reactions.\par
This article is organized as follows. Starting in Sec.~\ref{sect_discussion} with a brief discussion of our approach and relating it to other methods used in the literature, we proceed in Sec.~\ref{sect_QCDfactor} to formulate the collinear factorization of correlation functions in position space, including one-loop results for the investigated current combinations. In Sec.~\ref{sect_lattice} we detail the methods used in our lattice computation. We present our results in Sec.~\ref{sect_results}, before we conclude.
\section{Heuristic discussion}\label{sect_discussion}%
Here, we discuss a simple example for how the information on parton distributions at lightlike separations can be extracted from the study of Euclidean correlation functions. We start from the pion transition form factor $F_{\pi\gamma\gamma}(q_1^2,q_2^2)$ of the reaction $\pi^0(p) \to \gamma^{\ast}(q_1) + \gamma^{\ast}(q_2)$, which can be obtained from the matrix element of the product of two electromagnetic currents,%
\begin{align}
\MoveEqLeft\int\! d^4z\, e^{ i(q_1-q_2) \cdot z/2} \langle 0 | T\{j_\mu(\tfrac{z}{2}) j_\nu(-\tfrac{z}{2})\}| \pi^0(p) \rangle  
\notag\\
&= ie^2 \epsilon_{\mu\nu\alpha\beta} q_1^\alpha q_2^\beta F_{\pi\gamma\gamma}(q_1^2,q_2^2)\,, 
\label{eq:pi-gamma-gamma}
\end{align}%
where $e$ is the electric charge and $p = q_1 + q_2$ is the pion momentum. The form factor $F_{\pi\gamma\gamma}(q_1^2,q_2^2)$ can be measured experimentally, at least in principle. If at least one of the photon virtualities is large, the form factor can also be calculated in QCD in terms of a single nonperturbative function describing the quark momentum fraction distribution $u$ in the pion at small transverse separation, the pion DA. For the heuristic discussion in this section, we consider the leading contribution shown in Fig.~\ref{fig-pigammagamma}; the corrections are discussed in the next section. To this accuracy, one obtains~\cite{Lepage:1980fj}%
\begin{align}
  F_{\pi\gamma\gamma}(q_1^2,q_2^2) &= - \frac{2}{3} F_\pi \int_0^1\!\frac{du\,\phi_\pi(u)}{u q_1^2 + (1-u) q_2^2}\,,
\end{align}%
where $F_\pi\simeq\unit{92}{\mega\electronvolt}$ is the pion decay constant. If the form factor is measured for a wide range of photon virtualities, the pion DA $\phi_\pi(u)$ can be extracted from this relation (up to various higher order correction terms). In practice, such measurements are very difficult and experimental information is only available for kinematical situations where one virtuality is large and the second is close to zero~\cite{Aubert:2009mc,Uehara:2012ag}, which is not sufficient to map out the complete $u$-dependence.\par%
The integral $\int \! d^4z$ of \eqref{eq:pi-gamma-gamma} receives contributions from both spacelike and timelike separations. Spacelike correlation functions can readily be accessed in lattice simulations. However, addressing timelike distances is not at all straightforward. The central observation at the root of the recent development is that timelike contributions are not needed (in the present context) as the complete information on the pion DA in principle is already contained in the spacelike correlator.\par%
Indeed, to the same accuracy as above,%
\begin{align}
\langle 0 | T\{j_\mu(\tfrac{z}{2}) j_\nu(-\tfrac{z}{2})\}| \pi^0(p)\rangle 
& = \frac{2i\, F_\pi}{3 \pi^2 z^4} \epsilon_{\mu\nu\alpha\beta} p^\alpha z^\beta \Phi_\pi(p\cdot z)\,,  
\label{eq:pigg-in-z-space}
\end{align}%
where%
\begin{align}%
\Phi_\pi(p\cdot z) &= \int_0^1\!du\, e^{ i(u-1/2)p\cdot z} \phi_\pi(u)\,
\label{Phi_pi}
\end{align}%
is the pion DA in longitudinal position space, which is analogous to the Ioffe-time parton distribution in deep-inelastic lepton-hadron scattering~\cite{Ioffe:1969kf,Braun:1994jq}. The correlation function in Eq.~(\ref{eq:pigg-in-z-space}) can be calculated on the lattice for spacelike separations $z^2<0$ and in principle arbitrarily large values of the scalar product $p\cdot z$. In this way, $\Phi_\pi(p\cdot z)$ can be directly measured~\cite{Braun:2007wv}.\par%
Before going into details, we discuss the structure of the position space DA at a qualitative level to understand what kind of information can be obtained from such a measurement. Note that in the limit of exact isospin symmetry the equality $\phi_\pi(u) = \phi_\pi(1-u)$ holds. As a consequence $\Phi_\pi(p\cdot z)$ is a real function, \mbox{$\Phi_\pi(p\cdot z) = \Phi_\pi(-p\cdot z)$}, with the normalization condition $\Phi_\pi(0)=1$. The second derivative at the origin, $\Phi''_\pi(0)$, is related to the first nontrivial moment of $\phi_\pi(u)$, which is usually denoted as $\langle \xi^2\rangle$ and referred to as the second Mellin moment in the DA literature,%
\begin{align}
 \Phi''_\pi(0) = - \frac14\int_0^1\!du\, (2u-1)^2 \phi_\pi(u) \equiv - \frac14 \langle \xi^2\rangle\,,
\end{align}%
where $\xi=2u-1$. This moment can be obtained on the lattice using conventional techniques~\cite{Braun:2006dg,Arthur:2010xf,Braun:2015axa,Bali:2017ude} as the matrix element of a local operator that contains two covariant derivatives. Higher derivatives of $\Phi_{\pi}$ at the origin are sensitive to higher moments. It has become standard to write the pion DA as a series expansion in orthogonal (Gegenbauer) polynomials,%
\begin{align} \label{expansion_Gegenbauer}
\phi_\pi(u,\mu) &= 6u(1-u) \sum\limits_{n=0}^{\infty} a^\pi_n(\mu) C^{3/2}_n(2u-1) \,,
\end{align}%
where $a_0^\pi =1$. Note that to one-loop accuracy the coefficients $a^\pi_n (\mu)$ do not mix under evolution of the scale~$\mu$. Moments of the DA can be written in terms of the coefficients in the Gegenbauer expansion, e.g.,%
\begin{align}
  \langle \xi^2\rangle = \frac15 + \frac{12}{35} a^\pi_2\,.  
\end{align}%
\begin{figure}[t]%
\centering
\includegraphics[width=\columnwidth,clip]{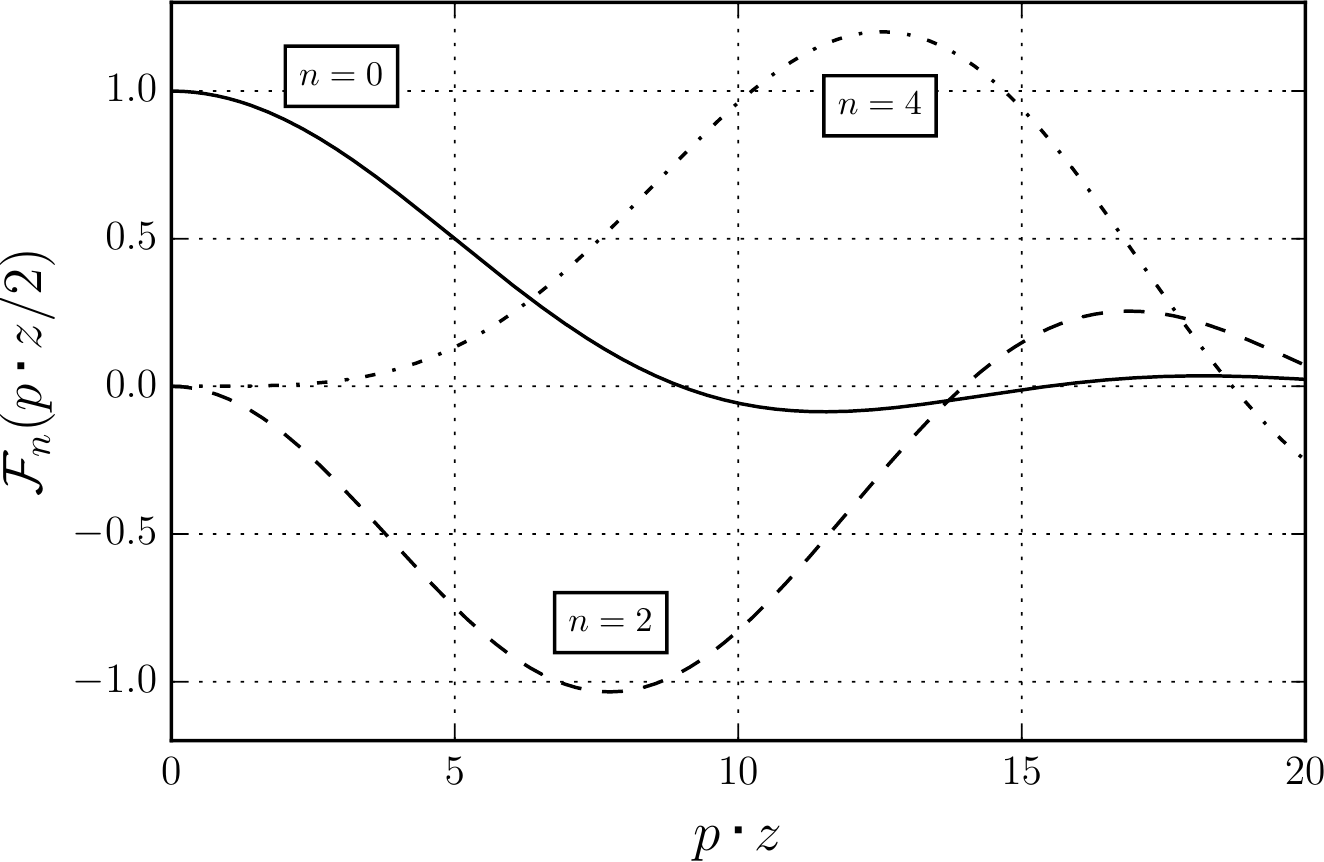}%
\caption{\label{fig-CPW}The first three conformal partial waves \eqref{calligraphicFn} in the expansion \eqref{expansion_CPW} of the pion DA in position space.}
\end{figure}%
The corresponding expansion of the DA in position space is in terms of Bessel functions (conformal partial waves~\cite{Braun:2007wv})%
\begin{align}\label{expansion_CPW}
\Phi_\pi(p\cdot z,\mu) &= \sum\limits_{n=0}^{\infty} a^\pi_n(\mu) \mathcal F_n(p\cdot z/2)\,,
\end{align}%
where
\begin{align}\label{calligraphicFn}
\mathcal F_n(\rho) &= \frac34 i^n \sqrt{2 \pi} (n+1) (n+2) \rho^{-\frac32} J_{n+\frac32}(\rho) \,.
\end{align}%
The first few conformal partial waves $\mathcal F_n(p\cdot z/2)$, $n=0,2,4$, are shown in Fig.~\ref{fig-CPW}. Since $\mathcal F_n(\rho) \sim \rho^n$ for $\rho\to 0$, the sum in (\ref{expansion_CPW}) for fixed $p\cdot z$ is converging very rapidly; only the first few Gegenbauer moments give a sizeable contribution. Conversely, this means that, aiming to extract the information on the pion DA beyond the first few moments, one has to include measurements at large~$p\cdot z$~\cite{Braun:2007wv}.\par%
To view this from a somewhat different perspective, consider, for illustrative purposes, the one-parameter class of models%
\begin{align} \label{DA_alpha_parametrization}
\phi_\pi^{(\alpha)} (u) &= \frac{\Gamma(2(\alpha+1))}{\Gamma(\alpha+1)^2} \bigl[u(1-u)\bigr]^\alpha\,,
\end{align}%
at the reference scale $\mu_0=\unit{2}{\giga\electronvolt}$. Three particular choices,
\begin{align}
\label{models}
\phi_\pi^{(1)} (u) &= 6 u(1-u)\,,
\notag\\*
\phi_\pi^{(1/2)} (u) &=  \frac{8}{\pi}\sqrt{u(1-u)}\,,
\notag\\*
\phi_\pi^{(0)} (u) &=  1\,,
\end{align}%
\begin{figure}[t]
\centering%
\includegraphics[width=\columnwidth,clip]{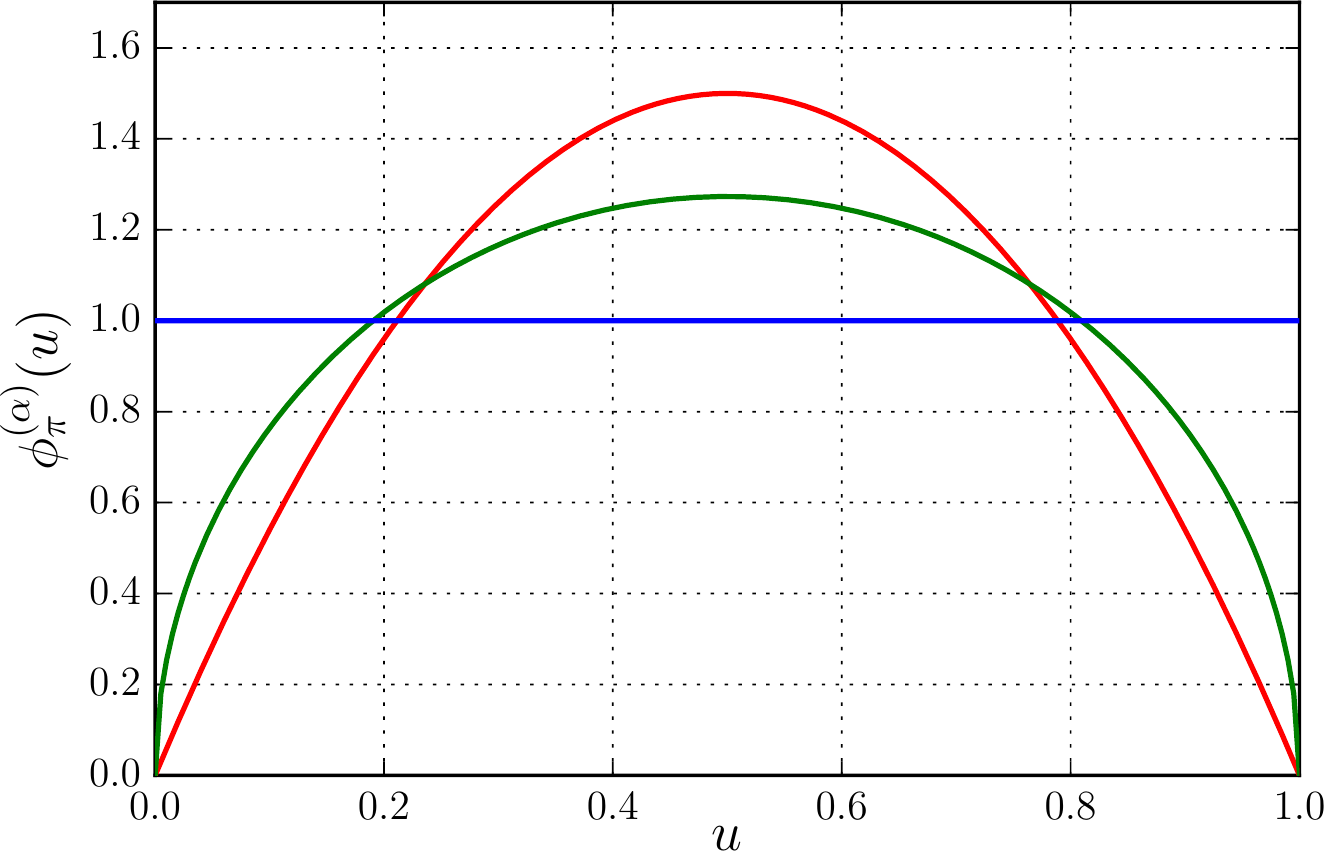}\\[0.3cm]
\includegraphics[width=\columnwidth,clip]{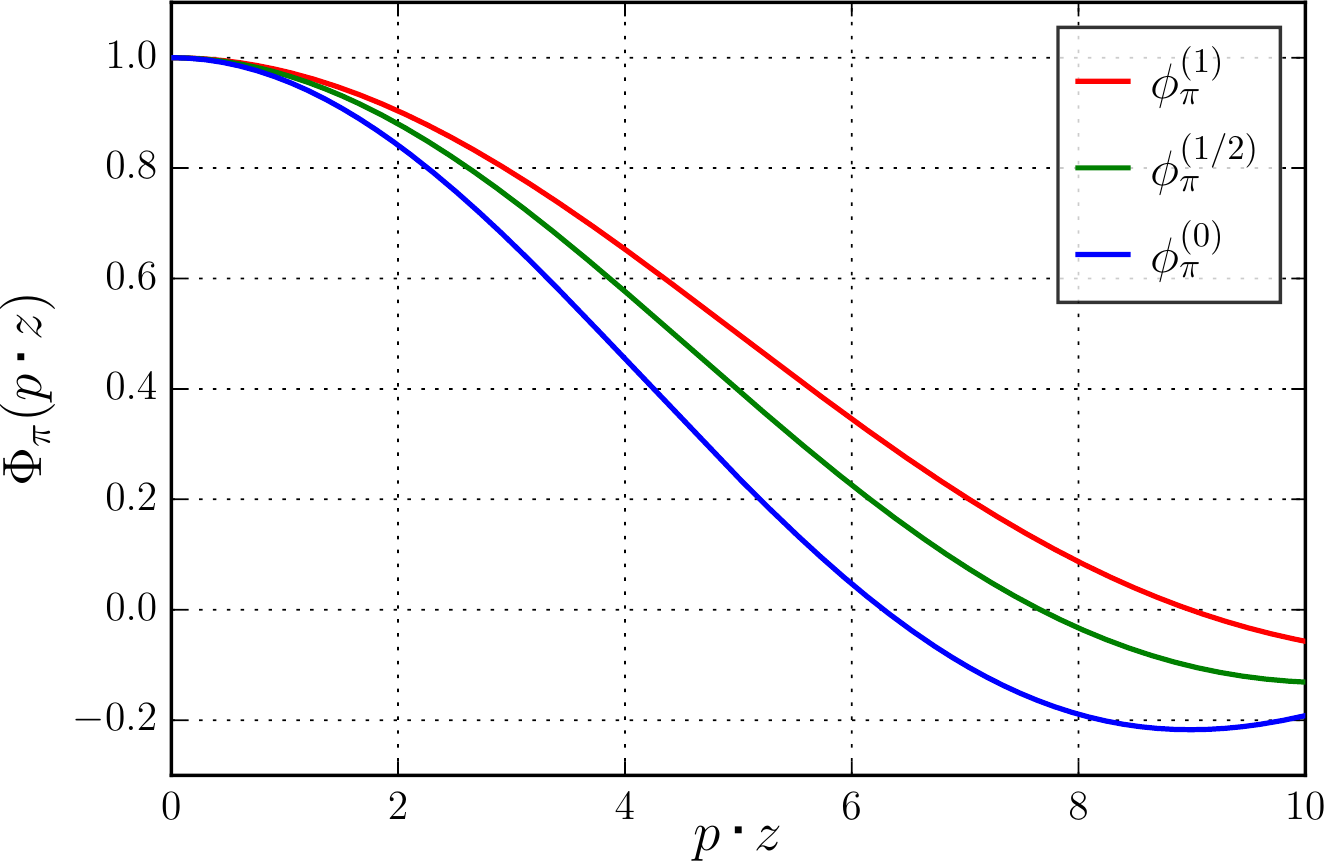}%
\caption{\label{fig:pionDAmodels}Three models for the pion distribution amplitude~\eqref{models} in momentum fraction (upper panel) and position space (lower panel). Note that $\Phi_{\pi}(-p\cdot z)=\Phi_{\pi}(p\cdot z)$.}
\end{figure}%
cover a wide range of shapes that appear to be phenomenologically acceptable. These longitudinal momentum fraction space DAs and the corresponding position space DAs $\Phi_\pi(p\cdot z)$ are plotted in Fig.~\ref{fig:pionDAmodels}. The differences between the models increase with $p\cdot z$. However, as demonstrated in Fig.~\ref{fig-CPW}, in the range accessible with present-day lattice calculations ($|p\cdot z| \lesssim 5$), the differences are almost entirely due to the variation of the second Gegenbauer moment: $a^\pi_2(\mu_0) = 0.389, 0.146, 0$ for the three above models, respectively.\par%
So far, we have discussed the situation at tree level. Taking into account QCD corrections, the position space pion DA $\Phi_\pi(p\cdot z)$ in Eq.~\eqref{eq:pigg-in-z-space} will be substituted by a function of both scalar invariants, $z^2$ and $p\cdot z$, of the form%
\begin{align} \label{twist_decomposition_heuristic}
\Phi^{\rm VV}_\pi(p\cdot z, z^2) &= C^{\rm VV}_2 (p\cdot z,z^2, u,\mu_F)\otimes \phi^{(2)}_\pi(u,\mu_F) \notag\\ 
&~ + z^2 C^{\rm VV}_4 (p\cdot z,z^2,u,\mu_F)\otimes \phi^{(4)}_\pi(u,\mu_F) \notag\\
&~+\mathcal{O}(z^4)\,,
\end{align}%
where $\phi^{(2)}_\pi \equiv \phi_\pi$ is the twist-$2$ DA. The $C^{\rm VV}_n$ are coefficient functions that depend at most logarithmically on $z^2$ and are calculable in perturbation theory, while $\mu_F$ is the factorization scale. We will tacitly assume using dimensional regularization and the modified minimal subtraction ($\MSbar$) scheme. The superscript VV indicates the dependence of the coefficient functions on the choice of the correlation function used to define the position space DA --- two vector currents for the present example, Eq.~\eqref{eq:pigg-in-z-space}. The leading- (and higher-)twist pion DAs are universal nonperturbative functions and independent of this choice. The power-suppressed $\mathcal{O}(z^2)$ correction terms correspond to higher-twist pion DAs, like~$\phi_\pi^{(4)}$. The factorization scale $\mu_F$ should be chosen similar in size to $2/\sqrt{-z^2}$ to prevent large logarithms from appearing in the coefficient functions.\par%
The function $\Phi^{\rm VV}_\pi(p\cdot z,z^2)$, and/or similar correlation functions with different choices of currents, can be calculated on the lattice within certain ranges of the two arguments. Different strategies have been suggested as to how useful information can be extracted from such lattice data. In this work, we follow the proposal of Ref.~\cite{Braun:2007wv} as well as our work~\cite{Bali:2017gfr}, carrying out the complete analysis in position space. We keep the distance between the currents sufficiently small to suppress higher-twist effects and to enable the perturbative evaluation of the coefficient functions at the scale $\mu_F\sim 2/\sqrt{-z^2}\geq\unit{1}{\giga\electronvolt}$, i.e., $\sqrt{-z^2}\lesssim\unit{0.4}{\femto\meter}$. At the same time $\sqrt{-z^2}$ should be much larger than the lattice spacing, in this work $a\approx\unit{0.071}{\femto\meter}$, to tame discretization effects.\par%
In the literature, it has also been suggested to carry out a one-dimensional Fourier transformation of the lattice data in order to define new observables that are closer in spirit to the initial DA in longitudinal momentum fraction space, e.g., a \emph{quasidistribution}~\cite{Lin:2014zya,Alexandrou:2015rja,Chen:2016utp,Alexandrou:2016jqi,Lin:2017ani,Zhang:2017bzy,Chen:2017gck,Chen:2017mzz,Alexandrou:2017huk,Alexandrou:2018pbm,Chen:2018xof,Chen:2018fwa},%
\begin{align}
   \phi^{\rm qu}_\pi(w) \sim \int_0^\infty \frac{d\lambda}{2\pi}\, e^{-i(w-1/2)\lambda p\cdot z} \Phi^{\rm XY}_\pi(\lambda p\cdot z,\lambda^2 z^2)\,, 
\end{align}%
or a \emph{pseudodistribution}~\cite{Radyushkin:2017cyf,Orginos:2017kos,Karpie:2017bzm},%
\begin{align}
   \phi^{\rm ps}_\pi(w) \sim \int_0^\infty  \frac{d\lambda}{2\pi}\, e^{-i(w-1/2)\lambda p\cdot z} \Phi^{\rm XY}_\pi(\lambda p\cdot z, z^2 )\,. 
\end{align}%
Both expressions are designed in such a way that, to leading-twist accuracy, they reproduce the pion DA at tree level. In the existing calculations which employ the above methods, two spatially separated quark fields are connected with a Wilson line. Equivalently, this construction can be viewed as a correlation function involving two bilinear currents with an auxiliary ``heavy'' quark field~\cite{Craigie:1980qs,Dorn:1986dt,Green:2017xeu} rather than the light quark field we use in Eq.~\eqref{eq:pigg-in-z-space}. Apart from employing a Wilson line~\cite{Chu:1990ps,Lin:2014zya,Alexandrou:2015rja,Chen:2016utp,Alexandrou:2016jqi,Lin:2017ani,Zhang:2017bzy,Chen:2017gck,Chen:2017mzz,Radyushkin:2017cyf,Orginos:2017kos,Karpie:2017bzm,Alexandrou:2017huk,Alexandrou:2018pbm,Chen:2018xof,Chen:2018fwa} or an auxiliary light quark propagator~\cite{Wilcox:1986ge,Braun:2007wv,Bali:2017gfr}, other obvious choices for connecting the two positions include a scalar propagator~\cite{Aglietti:1998ur,Abada:2001if} or a heavy quark propagator~\cite{Detmold:2005gg} or just employing Coulomb gauge~\cite{Gupta:1993vp}.\par
Another technical difference of the quasidistribution work relative to our approach is the use of the large-momentum factorization scheme at an intermediate step (large momentum effective theory~\cite{Ji:2014gla,Izubuchi:2018srq}) to emphasize that, for a large pion momentum and at a fixed quark momentum fraction, large-distance (i.e., higher-twist) contributions are suppressed.\par%
\section{QCD factorization}\label{sect_QCDfactor}%
\subsection{Collinear factorization in position space}\label{sect_QCDfactor_gen}%
A general approach to implement collinear factorization of QCD amplitudes in position space is provided by the light-ray operator product expansion (OPE)~\cite{Anikin:1978tj,Anikin:1979kq,Balitsky:1987bk,Mueller:1998fv,Balitsky:1990ck,Geyer:1999uq}. For a generic current product, one writes%
\begin{align}
\label{light-ray-OPE}
 J_1(z_1) J_2(z_2) &= Z_1 Z_2 \int\limits_0^1\!d\alpha\!\int\limits_0^1\!d\beta \, C_{12}(z_{12},\alpha,\beta,\mu_F)
\notag\\&~\times \Pi_{\rm l.t.}^{\mu_F}[\bar q (z_{12}^{(\alpha)}) \slashed{z}_{12} \gamma_5 q(z_{21}^{(\beta)})] + \ldots \,,
\end{align}%
where%
\begin{align}
 z_{12} = z_1-z_2\,, \qquad z_{12}^{(\alpha)} = (1-\alpha)z_1 + \alpha z_2\,,
\end{align}%
while $Z_k$ are the renormalization factors for the currents, $\Pi_{\rm l.t.}^{\mu_F}[\ldots]$ is the leading-twist projection operator, $C_{12}$ is the coefficient function, and the ellipses stand for higher-twist contributions. For simplicity, we disregard the flavor structure, showing only the contribution of flavor-nonsinglet axialvector operators that will be important for this work. The corresponding expression for the product of quark and antiquark fields connected by a Wilson line is exactly the same, with $Z_1,Z_2$ substituted by the quark field renormalization factors in spacelike axial gauge.\par%
The leading-twist projection of a nonlocal quark-antiquark operator is defined as the generating function of \emph{renormalized} local leading-twist operators (traceless and symmetrized over all indices), e.g.,%
\begin{align}\label{eq:LTprojector}
& \hspace*{-1cm}\Pi_{\rm l.t.}^{\mu_F}[\bar q (z_1) \slashed{z}_{12}\gamma_5 q(z_2)] \notag\\
&= \sum_{n=1}^\infty \sum_{k=0}^{n-1} \frac{z_{12}^{\mu_1}\ldots z_{12}^{\mu_n}  (-1)^k}{2^{n-1} k!(n-k-1)!} O_{\mu_1\ldots\mu_n}^{n,k}(z)\,,
\end{align}%
where $z=(z_1+z_2)/2$ and%
\begin{align}
  O_{\mu_1\ldots\mu_n}^{n,k}(z) &= \bar q(z)\gamma_{(\mu_1}\!\!\stackrel{\leftarrow}{D}_{\mu_2}\!\!\mydots \!\stackrel{\leftarrow}{D}_{\mu_{k+1}} \stackrel{\rightarrow}{D}_{\mu_{k+2}}\!\!\mydots 
\!\stackrel{\rightarrow}{D}_{\mu_{n})}\!\gamma_5 q(z)\,.
\end{align}%
Here and below, we indicate trace subtraction and symmetrization by enclosing the involved Lorentz indices in parentheses, for example $O_{(\mu\nu)} = \frac12 (O_{\mu\nu}+ O_{\nu\mu}) - \frac14 g_{\mu\nu} O^{\lambda}_{~\lambda}$.\par%
The light-ray OPE differs from the usual Wilson expansion in local operators by imposing a different power counting. In the latter case, one assumes that the distance between the currents is small, $|z_{12}| \sim \eta \Lambda^{-1}_{\rm QCD}$ with $\eta\to 0$, and the operator matrix elements are of order unity in this limit, $\langle O_{\mu_1\ldots\mu_n}^{n,k} \rangle \sim \Lambda^{n}_{\rm QCD}$. In this case, only a finite number of local operators on the r.h.s.\ of Eq.~\eqref{eq:LTprojector}
has to be kept, and also the higher-twist operators must be added progressing to higher powers of $\eta$: The relevant expansion parameter is the operator dimension, not the twist. The light-ray OPE assumes instead that $\langle O_{\mu_1\ldots\mu_n}^{n,k} \rangle \sim \eta^{-n} \Lambda^{n}_{\rm QCD}$ so that $z^{\mu_1}_{12}\ldots z^{\mu_n}_{12} \langle O_{\mu_1\ldots\mu_n}^{n,k} \rangle  = \mathcal{O}(1)$, and in this case, the series~\eqref{eq:LTprojector} must be resummed to all orders. Such a situation occurs if the hadron has large momentum, $|p| = \mathcal{O}(\eta^{-1})$ and hence $p\cdot z_{12} =  \mathcal{O}(1)$, since for generic hadronic matrix elements%
\begin{align}
 \langle H'(p) | O_{\mu_1\ldots\mu_n}^{n,k} | H(p)\rangle \sim p_{(\mu_1}\ldots p_{\mu_n)} \langle\!\langle O^{n,k} \rangle\!\rangle\,, 
\end{align}%
where the reduced matrix element $\langle\!\langle O^{n,k} \rangle\!\rangle = \mathcal{O}(1)$. Higher-twist operators of the same dimension have smaller spin (by definition). As a consequence, their matrix elements involve lower powers of the large momentum and are suppressed. At the amplitude level, expanding in powers of the large momentum corresponds to the classification in terms of the so-called collinear twist; see, e.g., Refs.~\cite{Geyer:1999uq,Geyer:2000ig}.\par%
Note that the above power counting is applicable both in Minkowski and Euclidean space. In Minkowski space, one can employ a reference frame where all components of the momentum are small and simultaneously the separation between the currents is almost lightlike, $|z_\mu| = \mathcal{O}(1)$, $z^2 = \mathcal{O}(\eta^2) \to 0 $. In this way, the usual interpretation as the light-cone expansion arises.\par%
The light-ray OPE provides a technique to deal with leading-twist projected operators~\eqref{eq:LTprojector} as a whole, avoiding the local expansion. These can be viewed as analytic operator functions of the separation between the currents (all short-distance and light-cone singularities are subtracted) and satisfy the equation~\cite{Balitsky:1987bk}%
\begin{align}
 \mbox{\Large $\Box$}_{z_{12}} \Pi_{\rm l.t.}^{\mu_F}[\bar q (z_1) \slashed{z}_{12}\gamma_5 q(z_2)] =0\,.
\end{align}%
Explicit expressions for the projection operator $\Pi_{\rm l.t.}^{\mu_F}$ can be found in Refs.~\cite{Balitsky:1987bk,Geyer:1999uq,Geyer:2000ig,Braun:2011dg}. This technique combined with the background field method has proven to be very efficient and has found many applications, e.g., in light-cone sum rules~\cite{Balitsky:1989ry} for the calculation of higher-twist contributions and for the derivation of the evolution equations for off-forward parton distributions~\cite{Belitsky:1999hf,Belitsky:1998gc}.\par%
Hadronic matrix elements of the operator~\eqref{eq:LTprojector} define leading-twist parton distributions. Specializing to our case, the pion DA is defined via%
\begin{align}
\begin{split}
\MoveEqLeft\langle 0 | \Pi_{\rm l.t.}^{\mu_F}[\bar q (\tfrac{z}{2}) \slashed{z}\gamma_5 q(- \tfrac{z}{2})] | \pi^0(p) \rangle \\
&= i F_\pi    \int_0^1\!du\, \Pi_{\rm l.t.}[(p\cdot z) e^{ i(u-1/2)p\cdot z}] \phi_\pi(u,\mu_F)\,, 
\end{split}
\end{align}%
where~\cite{Balitsky:1990ck}%
\begin{align}\label{Pi-exp}
\begin{split}
\MoveEqLeft\Pi_{\rm l.t.}[(p\cdot z) e^{ i(u-1/2)p\cdot z}]\\ 
&= \Big[(p\cdot z) - \frac{i}{8} (2u-1) m_\pi^2 z^2\Big] e^{ i(u-1/2)p\cdot z} + \mathcal{O}(z^4)\,. 
\end{split}
\end{align}%
The second term in the last line is the (twist-$4$) pion mass correction, which is analogous to the Nachtmann target mass correction in deep-inelastic scattering.%
\subsection{Choice of currents and one-loop results}\label{sect_QCDfactor_res}%
In this work, we perform a lattice study of the set of correlation functions%
\begin{align} \label{def:T_XY}
 \mathbb{T}_{\rm XY}(p \cdot z,z^2) = 
  \langle 0 | J^\dagger_{\rm X}(\tfrac{z}{2}) J_{\rm Y}(-\tfrac{z}{2})| \pi^0( p ) \rangle \ ,
\end{align}%
where the currents $J_{\rm X} \equiv \bar q\, \Gamma_{\rm X} u$ are defined as%
\begin{align}
J_{\rm S} &= \bar q\,  u\,, & J_{\rm P} &= \bar q \gamma_5 u\,,  \notag\\
J_{\rm V}^\mu &= \bar q\gamma^\mu u  \equiv J_{\rm V^\mu} \,, & J_{\rm A}^\mu &= \bar q\gamma^\mu\gamma_5 u  \equiv J_{\rm A^\mu} \label{def_dirac_structure}
\end{align}%
and contain an up quark $u$ and an auxiliary quark field~$q$. In this study, we assume that the auxiliary quark has different flavor than ($q\neq u,d$), but the same mass ($m_q=m_u$) as the light quarks. For convenience and better readability, we invoke the obvious notation $\mathbb{T}_{\rm VA}^{\mu\nu} \equiv \mathbb{T}_{\rm V^\mu A^\nu}$, etc.\par%
We do not consider the correlation functions of $\rm S$($\rm P$) with $\rm V$($\rm A$) currents because they are dominated by (chiral odd) higher-twist DAs. For the correlators with two Lorentz indices the most general invariant decomposition reads%
\begin{subequations}\label{decomposition}
\begin{align} 
\MoveEqLeft[1] \mathbb{T}_{\rm V V}^{\mu\nu}  = \frac{i \varepsilon^{\mu\nu\rho\sigma} p_\rho z_\sigma}{p\cdot z}  T_{\rm VV} \,, \qquad
\mathbb{T}_{\rm A A}^{\mu\nu}  = \frac{i \varepsilon^{\mu\nu\rho\sigma} p_\rho z_\sigma}{p\cdot z}  T_{\rm AA} \,, \\
\MoveEqLeft[1] 
\mathbb{T}_{\rm V A}^{\mu\nu} = \frac{p^\mu z^\nu \!+\! z^\mu p^\nu\!-\!g^{\mu\nu} p \cdot z}{p\cdot z} T_{\rm VA}^{(1)}
 + \frac{p^\mu z^\nu \!-\! z^\mu p^\nu}{p\cdot z}  T_{\rm VA}^{(2)} \vphantom{\int\limits_{X_x}]} 
\nonumber\\
&+ \frac{2 z^\mu z^\nu \!-\!g^{\mu\nu} z^2}{z^2}T_{\rm VA}^{(3)}
 + \frac{2 p^\mu p^\nu \!-\! g^{\mu\nu} p^2}{p^2}T_{\rm VA}^{(4)}
 + g^{\mu\nu}T_{\rm VA}^{(5)} \,,
\end{align}%
\end{subequations}%
where the prefactors are by construction invariant under rescaling of $z$ and all invariant functions $T_{\rm XY} \equiv T_{\rm XY}(p\cdot z, z^2)$ have the same mass dimension. The Lorentz decomposition for $\mathbb{T}_{\rm A V}^{\mu\nu}$ is obtained from the one for $\mathbb{T}_{\rm VA}^{\mu\nu}$ by replacing $\rm V \leftrightarrow A$. One can show that $T_{\rm VA} \equiv T_{\rm VA}^{\smash{(1)}}$ is the only invariant function in the $\rm VA$ correlator that receives contributions from the leading-twist DA at leading order in perturbation theory, so that we only consider this structure in what follows. The projection needed to isolate it is specified in Appendix~\ref{sect_QCDfactor_pro}. Finally, $\mathbb{T}_{\rm SP}$ and $\mathbb{T}_{\rm PS}$ are scalar functions which we write below as~$T_{\rm SP}$ and~$T_{\rm PS}$, respectively, to unify the notation.\par%
Separating a common overall prefactor, it is convenient to write the correlation functions in the form%
\begin{align}
\label{T_XY_LO}
   T_{\rm XY}(p \cdot z,z^2) &= F_\pi \frac{p\cdot z}{2\pi^2 z^4}  \Phi^{\rm XY}_\pi(p\cdot z,z^2)\,, 
\end{align}%
where to tree-level accuracy and neglecting higher-twist corrections $\Phi^{\rm XY}_\pi(p\cdot z,z^2) =  \Phi_\pi(p\cdot z) $ is the position space pion DA of Eq.~\eqref{Phi_pi}. We further separate the leading-twist (LT) contribution from the higher-twist (HT) part,%
\begin{align}%
\Phi^{\rm XY}_\pi(p \cdot z,z^2) &= \Phi^{\rm XY}_{\pi, \rm LT}(p \cdot z,z^2) +  \Phi^{\rm XY}_{\pi, \rm HT}(p \cdot z,z^2) \,,
\end{align}%
where the higher-twist contributions are of $\mathcal{O}(z^2)$, cf.\ Eq.~\eqref{twist_decomposition_heuristic}. The calculation of the one-loop, i.e., $\mathcal{O}(\alpha_s)$, correction at leading twist is relatively straightforward. Using the Gegenbauer expansion of the pion DA, Eqs.~\eqref{expansion_Gegenbauer} and~\eqref{expansion_CPW}, the result can be written as%
\begin{align}\label{nsum}
 \Phi^{\rm XY}_{\pi, \rm LT} &= \sum\limits_{n=0}^{\infty} H^{\rm XY}_n(p\cdot z,\mu)\, a^\pi_n(\mu)\,.
\end{align}%
Setting the renormalization and factorization scales to the same value $\mu= \mu_F$, we obtain, to $\mathcal{O}(\alpha_s)$ accuracy,%
\begin{widetext}
\begin{subequations} \label{Hn}
\begin{align}
H^{\rm SP}_n = H^{\rm PS}_n &=  \biggl[ 1 + \frac{\alpha_s C_F}{4\pi}(7 \eta - 11)  \biggr]\, \mathcal F_n(\rho) -\frac{\alpha_s C_F }{\pi} \! \int \limits_0^1 \! ds\, 
  \mathcal F_n(s\rho )\, \Biggl\{ (\eta - 4) \frac{\sin(\bar s\rho)}{2 \rho} 
  + \biggl[ (\eta - 2) \frac{s}{\bar s} + \frac{\ln(\bar s)}{\bar s}\biggr]_{\mathrlap{+}} \cos(\bar s\rho) \Biggr\}\,, \\
H^{\rm VA}_n = H^{\rm AV}_n &=  \biggl[ 1 + \frac{\alpha_s C_F}{4\pi}(\eta - 5)  \biggr]\, \mathcal F_n(\rho) -\frac{\alpha_s C_F }{\pi} \! \int \limits_0^1 \! ds\, 
  \mathcal F_n(s\rho)\, \Biggl\{ (\eta - 2) \frac{\sin(\bar s\rho)}{2\rho} 
  + \biggl(\biggl[ (\eta - \tfrac12) \frac{s}{\bar s} + \frac{\ln(\bar s)}{\bar s}\biggr]_{\mathrlap{+}} -\frac{1}{2} \biggr) \cos(\bar s\rho) \Biggr\}\,, \\
H^{\rm VV}_n = H^{\rm AA}_n &= \biggl[ 1 + \frac{\alpha_s C_F}{4\pi}(\eta - 5)  \biggr]\, \mathcal F_n(\rho) -\frac{\alpha_s C_F }{\pi} \! \int \limits_0^1 \! ds\, 
  \mathcal F_n(s\rho)\, \Biggl\{ (\eta - 2) \frac{\sin(\bar s\rho)}{2\rho} 
  + \biggl[ (\eta - \tfrac12) \frac{s}{\bar s} + \frac{\ln(\bar s)}{\bar s}\biggr]_{\mathrlap{+}} \cos(\bar s\rho) \Biggr\}\,,
\end{align}%
\end{subequations}%
\end{widetext}%
where $\alpha_s= \alpha_s(\mu)$, the functions $\mathcal{F}_n$ are defined in Eq.~\eqref{calligraphicFn}, $\rho=p\cdot z/2$, $C_F=\frac43$, $\eta=1+2\gamma_E+\ln(-z^2\mu^2/4)$, $\bar s=1-s$. In the following, we will choose $\mu\equiv2/\sqrt{-z^2}$. The plus prescription is defined as usual:%
\begin{align}
 \int\limits_0^1 ds \; f(s) \bigl[ g(s) \bigr]_+ \equiv   \int\limits_0^1 ds\; \bigl[f(s)-f(1)\bigr]  g(s) \,.
\end{align}%
The sum in~\eqref{nsum} converges very rapidly since%
\begin{align}
  \mathcal F_n(\rho) &\stackrel{\rho\to 0}{\simeq} 
\frac38 i^n \left(\frac{\rho}{2}\right)^n \frac{\sqrt{\pi}(n\!+\!1)(n\!+\!2)}{\Gamma(n+5/2)}\,,
\end{align}%
cf.\ Fig.~\ref{fig-CPW}, so that for moderate $p\cdot z$ only the first few Gegenbauer moments give a sizeable contribution~\cite{Braun:2007wv}. The two-loop corrections $\mathcal{O}(\alpha_s^2)$ are known for the $\rm VV$~correlator~\cite{Braun:2007wv} but not for other cases, so we do not include them in this study.\par%
The leading $\mathcal{O}(z^2)$ higher-twist contribution can be estimated using models for the twist-$4$ pion DAs derived in Refs.~\cite{Braun:1989iv,Ball:2006wn}; see also Appendix~\ref{App:HT}. One obtains%
\vspace{0.05cm}\begin{align}
 \Phi^{\rm XY}_{\pi, {\rm HT}} &=  \frac{z^2}{4} \!\! \int \limits_0^1 \!\! du \, \cos[(u-\tfrac12) p\cdot z] f^{\rm XY}(u) + \mathcal O(z^4)\,,
\end{align}%
where
\begin{widetext}
\begin{subequations}\label{def_higher_twist}
\begin{align} 
 f^{\rm SP} = (f^{\rm PS})^\ast &= -20\, \delta^\pi_2 u^2\bar u^2 + m^2_\pi u\bar u + \frac{ m^2_\pi}{2} u^2\bar u^2 \Big[14u\bar u -5 + 6a_2^\pi (3-10u\bar u)\Big] + \frac{i m_\pi^2}{2 (p\cdot z)} \,, \\
 f^{\rm VA} = f^{\rm AV}  &= -\frac{20}{3} \delta^\pi_2 u \bar u (1-6u \bar u) + \frac{m^2_\pi}{12}  u \bar u (19-18a_2^\pi) + \frac{m^2_\pi}{2}  u^2\bar u^2 \Big[7 u\bar u -8 + 18a_2^\pi (2-5u\bar u)\Big]   \,, \displaybreak[3]\\
 f^{\rm VV} &=  \frac{80}{3} \delta^\pi_2 u^2\bar u^2 + \frac{m^2_\pi }{12} u^2\bar u^2 \Big[42 u\bar u -13 + 18a_2^\pi (7-30u\bar u)\Big]  \,, \\
 f^{\rm AA} &=  \frac{80}{3} \delta^\pi_2 u^2\bar u^2 + \frac{m^2_\pi }{12} u^2\bar u^2 \Big[42 u\bar u -13 + 18a_2^\pi (7-30u\bar u)\Big] + 2 m^2_\pi u\bar u   \,.
\end{align}%
\end{subequations}%
\end{widetext}%
The parameter $\delta^\pi_2$ is defined in Eq.~\eqref{def:delta-pi-2}. QCD sum rule estimates yield $\delta^\pi_2 \simeq \unit{0.2}{\giga\electronvolt\squared}$ at the scale \mbox{$\mu=\unit{1}{\giga\electronvolt}$~\cite{Novikov:1983jt,Braun:1989iv,Ball:2006wn}}. In comparison, the $\mathcal{O}(m_\pi^2)$ terms are rather small.\par%
For reasons that will be explained in Sec.~\ref{subsect_disc}, we choose to analyze linear combinations of correlation functions, $\rm PS+SP$, $\rm VA+AV$, and $\rm VV+AA$, to which the leading quark mass correction originating from the chiral odd part of the auxiliary quark propagator does not contribute. In this sum, e.g., the imaginary parts of the SP and PS correlators cancel each other and drop out. For the $\rm VA+AV$ case, the terms linear in the quark mass $m_q$ drop out completely after applying the projection onto the invariant function of interest as described in Appendix~\ref{sect_QCDfactor_pro}. Note that the entire difference between $f^{\rm VV}$ and $f^{\rm AA}$ is due to this quark mass correction, which is converted into an $\mathcal{O}(m_\pi^2)$ term using the axial Ward identity and $m_q=m_u$.\par%
As observed already in Ref.~\cite{Braun:2007wv}, the higher-twist correction in the VV channel has opposite sign compared to the leading-twist term. We find a similar behavior for the AV channel. In contrast, the twist-$4$ correction for the SP correlation function has the same sign as the leading-twist contribution. Numerically, the higher-twist corrections turn out to be approximately the same size as the leading perturbative correction at $\sqrt{-z^2}/2 \sim \unit{0.2}{\femto\meter} \simeq \unit{1}{\power{\giga\electronvolt}{-1}}$ and become gradually less important for smaller distances. At the lowest scale considered in this study, $\unit{1}{\giga\electronvolt}$, and at $p \cdot z=0$, the combined one-loop and higher-twist correction yields approximately $-40\%$, $-20\%$, and $+50\%$ for the $\rm VV$, $\rm VA$, and $\rm SP$ channels, respectively. We will find that these estimates are strongly supported by our lattice data, cf.\ Sec.~\ref{sect_results}.%
\section{Lattice calculation}\label{sect_lattice}%
We employ the same gauge ensemble as in Ref.~\cite{Bali:2017gfr} (ensemble~IV of Ref.~\cite{Bali:2014gha}, generated by the QCDSF and RQCD collaborations), which allows a direct comparison between the sequential source method~\cite{Martinelli:1988rr} (used in Ref.~\cite{Bali:2017gfr}) and the stochastic method (applied in this work) for the scalar-pseudoscalar channel. We employ the Wilson gluon action with two mass-degenerate flavors of nonperturbatively order $a$ improved Sheikholeslami--Wohlert~\cite{Sheikholeslami:1985ij} (i.e., clover) Wilson fermions. The lattice consists of $32^3\times64$ points with periodic boundary conditions (antiperiodic in time for the fermion fields). The inverse gauge coupling parameter reads $\beta=5.29$, and the hopping parameter value is $\kappa=0.13632$. This corresponds to the lattice spacing $a\approx\unit{0.071}{\femto\meter}=(\unit{2.76}{\giga\electronvolt})^{-1}$~\cite{Bali:2012qs} and a pion mass $m_\pi=0.10675(59)/a \approx \unit{295}{\mega\electronvolt}$~\cite{Bali:2014nma}. To reduce autocorrelations we have used a bin size $N_{\rm bin}=20$ for the $N_{\rm conf}=2000$ configurations we have analyzed, cf.\ Table~\ref{tab_momenta}. In order to improve the overlap between the interpolating current at the source and the pion state at large momentum, we employ the momentum smearing technique of Ref.~\cite{Bali:2016lva} (see also Ref.~\cite{Bali:2017ude}) with APE-smeared spatial gauge links~\cite{Falcioni:1984ei}.\par%
The operator renormalization is performed as described in Ref.~\cite{Gockeler:2010yr}: The renormalization factors are calculated nonperturbatively within the RI${}^\prime$-MOM scheme~\cite{Martinelli:1994ty,Chetyrkin:1999pq} (along with a subtraction of lattice artifacts in one-loop lattice perturbation theory). These are then converted to the $\MSbar$ scheme using three-loop (continuum) perturbation theory. The corresponding factors for the $\MSbar$ scale $\mu=\unit{2}{\giga\electronvolt}$ can be found in Table~III of Ref.~\cite{Bali:2014nma}. To be consistent, we employ the $N_f=2$ specific running of~$\alpha_s$ in all perturbative calculations. To this end, we combine the results of Refs.~\cite{Bali:2012qs} and~\cite{Fritzsch:2012wq} to obtain a value of $\alpha_s$ at $1000/a\approx\unit{2.76}{\tera\electronvolt}$. From there, we evolve it downward using five-loop running~\cite{Baikov:2016tgj}. The pseudoscalar and scalar currents are evolved to other scales using the four-loop mass anomalous dimension, which is consistent with the order used in Ref.~\cite{Gockeler:2010yr}. The numerical values of the $N_f=2$ coefficients are summarized, e.g., in Ref.~\cite{Baikov:2014qja}, which also includes the five-loop calculation.\par%
Disconnected quark line diagrams have proven to be notoriously challenging in lattice simulations. They can be avoided by implementing an appropriate flavor structure of our currents: we pretend that the auxiliary quark field $q$ of Eq.~\eqref{def:T_XY} is of a different flavor but shares its mass with the light quarks: $m_q=m_u=m_d$. This does not present any limitation as the perturbative matching is carried out using the same conventions.\par%
In the following sections, we use boldface letters for the space components of the distance and momentum, $(z^\mu) = (0,\mathbf{z})$ and $(p^\mu) = (E_{\mathbf p}, \mathbf{p})$. In the actual lattice calculation, we evaluate the three-point functions using currents positioned at $\mathbf z$, relative to our origin $\mathbf 0$. These are ``shifted'' afterward to the symmetric locations as in Eq.~\eqref{def:T_XY} by multiplication with the appropriate phase.\par%
\subsection{Stochastic estimation of correlation functions}\label{subsect_stochastic_estimation}%
\begin{figure}[tb]%
\centering%
\includegraphics[width=\columnwidth]{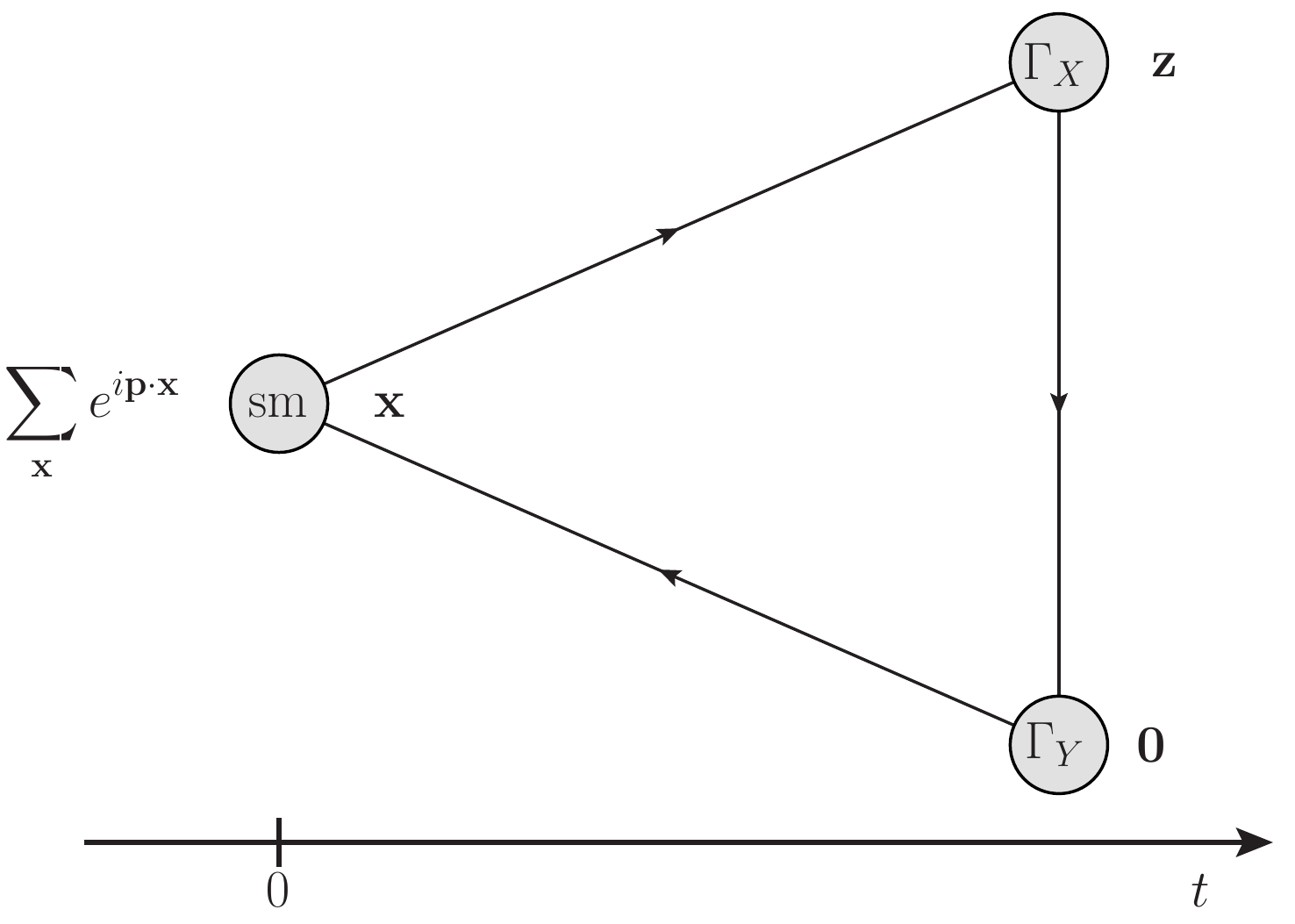}%
\caption{\label{fig_triangle}The relevant triangle diagram, with a smeared interpolating current for the pion at $t=0$. The Fourier transform corresponds to an incoming pion with momentum $\mathbf p$ (for $t>0$) or an outgoing pion with momentum $-\mathbf p$ (if $t<0$).}
\end{figure}%
We wish to compute the correlation functions, Eq.~\eqref{def:T_XY}. The corresponding three-point functions for different \mbox{$\Gamma$ structures} are depicted in Fig.~\ref{fig_triangle}, where the straight lines correspond to quark propagators. The momentum-smeared, momentum-projected pion source is located at the Euclidean time slice $0$. Translational invariance of the correlation function implies that $\mathbf{z}$ and~$\mathbf{0}$ can be shifted to the positions $\mathbf{z}/2$ and $-\mathbf{z}/2$, respectively, by multiplication with the phase $e^{i\mathbf{p}\cdot\mathbf{z}/2}$. Previously, in Ref.~\cite{Bali:2017gfr}, we computed propagators, starting from a point source at the position $(t,\mathbf{0})$, smeared the resulting propagator at the time slice $t=0$ and computed a sequential propagator~\cite{Martinelli:1988rr} from there. This propagator and the original propagator were then contracted with the $\Gamma_{\rm X}$ structure at $(t,\mathbf{z})$, making use of $\gamma_5$-Hermiticity of the propagator~$G$, i.e.,~$G_{xy}=\gamma_5 G_{yx}^{\dagger}\gamma_5$.\par%
In order to increase the statistics, ideally one would average over different spatial positions $\mathbf{y}$ of the current~$J_{\rm Y}$, placing~$J_{\rm X}$ at positions $\mathbf{y}+\mathbf{z}$, keeping the relative distance vector fixed. It turns out that this is indeed possible, introducing stochastic propagators~\cite{Bali:2009hu}, albeit at the cost of additional (but small) stochastic noise.\par%
Therefore, our new approach is to start from a momentum-smeared pion source at $t=t_{\rm src}$ and compute stochastic forward propagators from there, using the ``one-end trick''~\cite{Foster:1998vw}. As with the sequential method adopted by us previously, the external momentum is fixed at the source. Since we need to keep the distance $\mathbf{z}$ between the local currents $J_{\rm X}$ and $J_{\rm Y}$ at the sink fixed, volume averaging would not be possible if we created a sequential propagator at the sink. Instead, we use a second stochastic volume source at $t_{\rm sink}$ to connect these two currents. In order to reduce the associated stochastic noise we utilize the hopping parameter expansion in the way suggested in Refs.~\cite{Bali:2005pb,Bali:2005fu} (see also Refs.~\cite{Thron:1997iy,Michael:1999rs} for related work) to block out the dominant short-distance noise contributions when connecting the two currents with a stochastic propagator. Below we describe our implementation in detail.\par%
We define the momentum smearing operator $F_{\mathbf{p}}=\Phi_{(\zeta\mathbf{p})}^n$ with $n$ smearing iterations ($n=200$ in our calculation). This is diagonal in spin and constructed on the time slice $t_{\rm src}$, iteratively applying the operation%
\begin{align}
\label{Smear}
(\Phi_{(\mathbf{k})} q)_{\mathbf{x}} = \frac{1}{1+6\varepsilon} \left[q_{\mathbf{x}}+\varepsilon \smashoperator{\sum_{j=\pm 1}^{\pm 3}} U_{\mathbf{x},j}e^{-i\mathbf{k}\cdot\jhat} q_{\mathbf{x}+a\jhat}\right],
\end{align}%
where $U_{\mathbf{x},j}$ is an APE-smeared~\cite{Falcioni:1984ei} spatial gauge link connecting the lattice points $(t_{\rm src},\mathbf{x})$ and $(t_{\rm src},\mathbf{x}+a\jhat)$; for details, see Refs.~\cite{Bali:2016lva,Bali:2017ude}. In practice, this smearing is implemented by multiplying the spatial connectors within the time slice in question by the appropriate phases, $U_{\mathbf{x},j}\mapsto e^{-iak_j}U_{\mathbf{x},j}$, where $\mathbf{k}=\zeta\mathbf{p}$. We choose $\zeta=0.8$ and $\epsilon=0.25$.\par%
We write the Wilson--Dirac operator as%
\begin{align}
D=\frac{1}{2a\kappa}\left(\mathds{1}-H\right)\,.
\label{eq:dirac}
\end{align}%
We also define the time, spin, and color diagonal momentum projection operator $\varphi_{\mathbf{p}}$ with the components%
\begin{align}
(\varphi_{\mathbf p})_{xy}=e^{-i\mathbf{p}\cdot\mathbf{x}}\delta_{xy}\,.
\end{align}%
Note that $F_{\mathbf{p}}=F_{\mathbf{p}}^{\dagger}=F_{\mathbf{-p}}^{\intercal}$ is self-adjoint, while $\varphi_{\mathbf{p}}=\varphi_{\mathbf{p}}^{\intercal}=\varphi^{\dagger}_{-\mathbf{p}}$ and $D=\gamma_5 D^{\dagger}\gamma_5$ are not. We start from a stochastic $\mathbb{Z}_2\otimes i\mathbb{Z}_2$ wall source $\xi$ with $\xi_{(t,\mathbf{x})}^{i,\alpha}=(\pm 1\pm i)/\sqrt{2}\,\delta_{t\,t_{\rm src}}$, where $i$ and $\alpha$ denote the color and spin indices, respectively. We then solve%
\begin{align}
aD\, \chi &=F_{-\mathbf{p}}\varphi_{-\mathbf{p}}\xi\,,&
aD\, \widetilde{\chi} &=F_{\mathbf{p}}\xi\,,
\label{eq:chi}
\end{align}%
where $\chi$ and $\widetilde{\chi}$ (as well as $\xi$) are Dirac vectors with color, spin, and spacetime components. Above, we have suppressed these indices for enhanced readability. In our conventions $\xi$, $\chi$, and $\widetilde{\chi}$ (as well as $\eta$ and $s$, which will be introduced below) are dimensionless.\par%
We define a momentum-smeared interpolator that, when applied to the vacuum, will create states with the quantum numbers of a $\pi^0$ carrying the (spatial) momentum $\mathbf{p}$,%
\begin{align}
O_{\mathbf{p}}^{\dagger}(t)&=a^3\sum_{\mathbf{x}}e^{i\mathbf{p}\cdot\mathbf{x}}O_{\pi}^{\dagger}(x)\,,\nonumber\\
O_{\pi}^{\dagger}(x)&=[\bar u F_{-\mathbf{p}}]_x \gamma_5 [F_{\mathbf{p}}u]_x - (u \rightarrow d)\,,
\end{align}%
and a local isovector current $J^v= \bigl( \bar{u}\Gamma u - \bar{d}\Gamma d\bigr)/2$ with an arbitrary Dirac structure $\Gamma$. In the following we assume, for the sake of readability, that all sources have been shifted to $t_{\rm src}=0$ (exploiting translational invariance) and denote the source-sink distance as $t$. We can now obtain the average over the spatial volume $V_3$ of smeared-local two-point functions%
\begin{align}
C(\mathbf{p},t)&=\left\langle 0\left|J^v(t,{\bf 0})O^{\dagger}_{\mathbf{p}}(0)\right|0\right\rangle\nonumber\\
&=\frac{a^6}{V_3}
\sum_{\mathbf{x},\mathbf{y}} e^{i\mathbf{p}\cdot(\mathbf{x}-\mathbf{y})}
\left\langle 0\left|J^v(t,\mathbf{y})
O_{\pi}^{\dagger}(0,\mathbf{x})\right|0\right\rangle
\label{correlator}
\end{align}%
as an inner product over color, spin, and (three-dimensional) space:%
\begin{align} 
C(\mathbf{p},t)
&=\frac{-a^6}{2V_3}\langle \operatorname{tr}\gamma_5 F_{\mathbf{p}} \contraction{}{u}{{}_0}{\bar{u}} u_0 \bar{u}_t \Gamma\varphi_{\mathbf{p}}\contraction{}{u}{{}_t}{\bar{u}}  u_t\bar{u}_0F_{-\mathbf{p}}\varphi_{-\mathbf{p}}\rangle + (u \rightarrow d) \nonumber \\
&=\frac{-1}{a^2 V_3}\left\langle\left(\xi,\gamma_5 F_{\mathbf{p}}D^{-1}_{0t}\Gamma \varphi_{\mathbf{p}}D^{-1}_{t0}F_{-\mathbf{p}}\varphi_{-\mathbf{p}}\xi\right)\right\rangle\nonumber\\
&=\frac{-1}{V_3}\left\langle \left(\widetilde{\chi}_t,\varphi_{\mathbf{p}} \gamma_5 \Gamma \chi_t\right)\right\rangle\,.
\end{align}%
Here we have suppressed all unnecessary indices. The disconnected contractions drop out since we have exact isospin symmetry. The minus sign in the first line is due to fermion anticommutation. Within the scalar product $(A,B)=A^{\dagger}B$, we sum over all indices that are not displayed on either side, in this case spatial position, color, and spin. In the second line, we used $a^{-4}D^{-1}_{0t}=\contraction[.5ex]{}{q}{{}_0}{\bar{q}} q_0 \bar{q}_t$ for $q\in\{u,d\}$. In the last step, we made use of the orthonormality of the noise vectors when averaged $\sum_{XY}\langle \xi_{Y}^*A_{YX}\xi_{X}\rangle=\langle\sum_X A_{XX}\rangle$, where $X,Y$ represent multi-indices, and of the $\gamma_5$-Hermiticity of the propagator.\par%
Inserting a complete set of states in Eq.~\eqref{correlator} and choosing an axialvector current at the sink with $\Gamma=\gamma_0 \gamma_5$ gives%
\begin{align}
C^{\rm 2pt}(\mathbf{p},t)&=\sum_{n}
\langle 0|A^v_0(0)|n(p)\rangle
\frac{e^{-E_n(\mathbf{p})t}}{2E_n(\mathbf{p})}
\langle n(p)|O_{\pi}^{\dagger}(0)|0\rangle\nonumber\\
&\longrightarrow
Z_{\pi}(\mathbf{p})
\frac{e^{-E_{\pi}(\mathbf{p})t}}{2E_{\pi}(\mathbf{p})}\langle 0|A^v_0(0)|\pi^0(p)\rangle\quad (t\rightarrow\infty)\,.\label{eq:decom}
\end{align}%
The overlap factor $Z_{\pi}(\mathbf p)=\langle \pi^0(p) |O^\dagger_{\pi}(0)| 0 \rangle$ depends on the (momentum-smeared) interpolator, while%
\begin{align}
\langle 0|A^v_\mu(0)|\pi^0(p)\rangle=i F_{\pi} p_\mu
\end{align}%
defines the pion decay constant.\par%
To construct the desired three-point function, we use additional spin-partitioned~\cite{Bernardson:1993he} (also referred to as spin explicit or ``diluted'' in the literature) stochastic sources $\eta^{(k,\alpha)}$, $k=1,\dots,n_{\rm st}$, and $\alpha=1,\dots,4$, with the components $\eta_{(t,\mathbf{x})}^{(k,\alpha) i,\beta}=r_{\mathbf{x}}^{(k) i} \,\delta^{\alpha\beta}\,\delta_{t\,t_{\rm sink}}$. The random variables $\smash{r_{\mathbf{x}}^{(k) i}}$ take the values $(\pm 1\pm i)/\sqrt{2}$. We then solve%
\begin{align}
        aD \, s^{(k,\alpha)}=\eta^{(k,\alpha)}
\end{align}%
for each value of $k$ and $\alpha$ to obtain $s^{(k,\alpha)}$. The lattice propagator $G=a^{-4}D^{-1}$ from $(t,\mathbf{y})$ to $(t,\mathbf{y}+\mathbf{z})$ can now be estimated as%
\begin{align}
        G_{(t,\mathbf{y}+\mathbf{z})(t,\mathbf{y})}\approx \frac{1}{a^3 n_{\rm st}} \sum_{k,\alpha} s^{(k,\alpha)}_{(t,\mathbf{y}+\mathbf{z})}\eta^{(k,\alpha)\dagger}_{(t,\mathbf{y})} \,,
\end{align}%
up to a stochastic error that decreases $\propto 1/\sqrt{N_{\rm conf}n_{\rm st}}$, where $N_{\rm conf}$ is the number of gauge configurations and $n_{\rm st}=10$ is the number of spin-partitioned stochastic sources.\par%
The operator $H$ in Eq.~\eqref{eq:dirac} only couples nearest neighbors for the action we use. Employing the geometric series%
\begin{align}
a^3 G&=(aD)^{-1}=2\kappa\left(\mathds{1}-H\right)^{-1}
 =2\kappa\smash[b]{\smashoperator{\sum_{j\geq 0}}}H^j\nonumber\\
&=2\kappa\smashoperator{\sum_{j=0}^{m(\mathbf{z})-1}}H^j+2\kappa\smashoperator{\sum_{j\geq m(\mathbf{z})}}H^j\nonumber\\
&=2\kappa\smashoperator{\sum_{j=0}^{m(\mathbf{z})-1}}H^j+H^{m(\mathbf{z})}a^3G
\,,\label{eq:hpe}
\end{align}%
where%
\begin{align}
        m(\mathbf{z})=\sum_{i=1}^3\min\left(\frac{|z_i|}{a},\frac{L-|z_i|}{a}\right)\,,
\end{align}%
we can split up the propagator into the first sum in Eq.~\eqref{eq:hpe} that does not contribute at distances $\mathbf{z}$ (and distances that are separated by a larger number of hops) and a part that contributes. In the stochastic estimation, the first part still adds to the noise. This undesirable effect can be removed, left multiplying the solution with $H^{m(\mathbf z)}$~\cite{Bali:2005pb,Bali:2005fu}.\par%
Looping over momenta and times, we define temporary scalar fields%
\begin{align}
K_{\rm X}^{(m,k,\alpha)}(\mathbf{y})&=\widetilde{\chi}^{\dagger}_{(t,\mathbf{y})}\gamma_5\Gamma_{\rm X} H^ms^{(k,\alpha)}_{(t,\mathbf{y})}\,, \label{KX}\\
K_{\rm Y}^{(k,\alpha)}(\mathbf{y})&=e^{-i\mathbf{p}\cdot\mathbf{y}}\eta^{(k,\alpha)\dagger}_{(t,\mathbf{y})}\Gamma_{\rm Y}\chi_{(t,\mathbf{y})} \label{KY}
\end{align}%
for $m\leq 10$ and all currents of interest. These fields implicitly depend on $\mathbf{p}$ and $t$. Also note that the solutions $\chi$ and $\widetilde{\chi}$ of Eq.~\eqref{eq:chi} depend on the momentum $\mathbf{p}$. The three-point correlation functions can now readily be obtained by replacing $J^v(t,\mathbf{y}) \mapsto J^\dagger_{\rm X}(t,\mathbf z/2) J_{\rm Y}(t,-\mathbf z/2)$ in Eq.~\eqref{correlator} (cf.\ Fig.~\ref{fig_triangle}). The result reads%
\begin{align}
C^{\rm 3pt}_{\rm XY}(\mathbf{p},t,\mathbf{z})=\frac{-e^{\frac{i}{2}\mathbf{p}\cdot\mathbf{z}}}{a^3 V_3 n_{\rm st}}
\smash{\sum_{\mathbf{y},k,\alpha}}
\left\langle K_{\rm X}^{(m,k,\alpha)}(\mathbf{y}+\mathbf{z})K_{\rm Y}^{(k,\alpha)}(\mathbf{y})\right\rangle\,,
\end{align}%
where the value of $m\leq m(\mathbf{z})$ used within the set of precomputed fields $K_{\rm X}^{(m,k,\alpha)}$ is selected as large as possible for each distance. Note that we have already shifted the above correlation function to the symmetric position. In our study we limit ourselves to the range $|z_i|\leq 5a$.\par%
With the previous sequential source method, first one propagator (12 solves) had to be computed. Then for each additional momentum and time separation, two smearing operations were required as well as an additional propagator (12 solves). In our implementation of the new method, we vary the distance between the pion source and the sink by changing the time slice where the pion source is placed, enabling us to reuse the stochastic solutions $H^ms^{(k,\alpha)}$ and sources $\eta^{(k,\alpha)}$ of Eqs.~\eqref{KX} and~\eqref{KY}. This part requires $4n_{\rm st}=40$ solves with a minimal overhead from applying the hopping parameter expansion. For each momentum and time separation, a new pion source is seeded, necessitating only two additional smearing operations and two additional solves.\par%
In total, not even taking into account that there is an additional gain from the two possibilities of connecting the valence quark propagators with the stochastic propagator of the auxiliary field (giving us for each momentum $\mathbf{p}$ the momentum $-\mathbf{p}$ almost for free), the new method does not only allow for a volume average, thereby reducing statistical errors, but turns out to be cheaper by about a factor of two in terms of the total computer time.\par%
The three-point function $C^{\text{3pt}}_{\rm XY}$ admits the same spectral decomposition, Eq.~\eqref{eq:decom}, as the two-point function~$C^{\text{2pt}}$. Just the matrix element needs to be replaced: $\langle 0|A^v_0(0)|\pi^0(p)\rangle\mapsto \langle 0|J_{\rm X}^{\dagger}(0,\mathbf z/2)J_{\rm Y}(0,-\mathbf z/2)|\pi^0(p)\rangle$. The overlap factor $Z_{\pi}(\mathbf{p})$ and the exponential decay cancel when taking the ratio of these two functions. Therefore, in the limit of large Euclidean times, where excited state contributions are exponentially suppressed, the ratio can be related to the matrix element of interest,%
\begin{align}%
 \frac{\mathbb{T}_{\rm XY}(p\cdot z,z^2)}{F_\pi} &= \frac{Z_{\rm X}(\mu)Z_{\rm Y}(\mu)}{Z_{\rm A}} \frac{C^{\text{3pt}}_{\rm XY}(\mathbf p,t,\mathbf z)}{C^{\text{2pt}}(\mathbf p,t)} i E_{\pi}(\mathbf p) \,,\label{eq:rato}
\end{align}%
where $Z_{\rm X}$ is the renormalization factor of the local current $J_{\rm X}$ with respect to the $\MSbar$ scheme~\cite{Gockeler:2010yr}. For the scalar and the pseudoscalar currents, the renormalization factors acquire a scale dependence due to their anomalous dimension.\par%
\subsection{Reducing discretization effects}\label{subsect_disc}%
\begin{figure}[t]%	
\centering%
\includegraphics[width=\columnwidth]{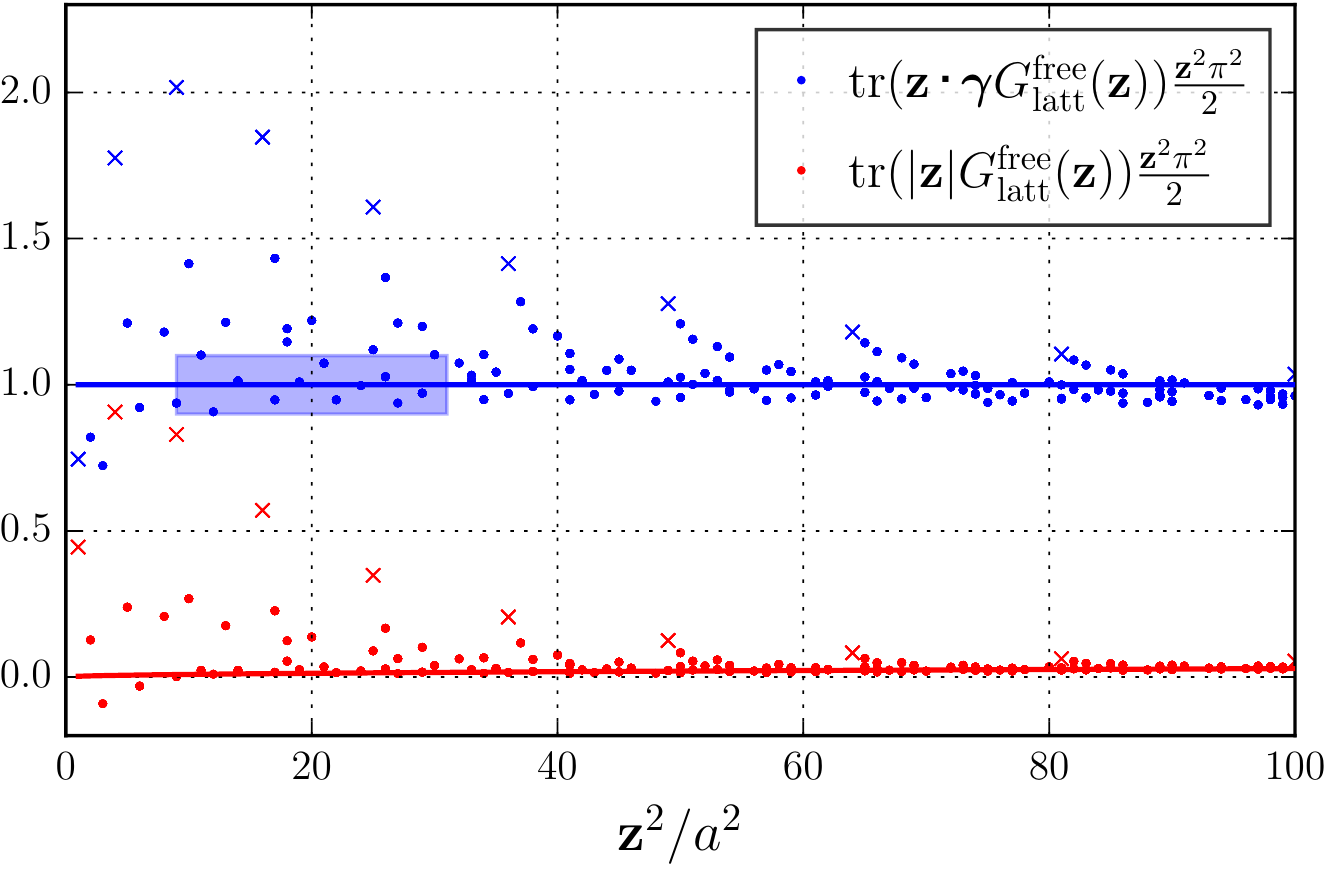}
\caption{\label{fig_disc}The (free field) discretization effects of the Wilson propagator compared to the continuum expectation (full lines) for the different Dirac structures. The points marked with a cross correspond to directions along a lattice axis. The shaded area marks the distances that are actually used in the analysis, where the upper limit comes from the constraint $\mu=2/|\mathbf z|>\unit{1}{\giga\electronvolt}$. It is clear that smaller lattice spacings will improve the situation considerably.}%
\end{figure}

In the continuum, the \emph{chiral even} part~($\propto \slashed z$) of the propagator connecting the two local currents gives the most important contribution, while the \emph{chiral odd} part~($\propto \mathds 1$) is suppressed by a factor $m \sqrt{-z^2}$ and, thus, can be set to zero in a first approximation. However, with Wilson fermions the situation is completely different. We find that the contribution from the chiral odd part, which suppresses the doublers and breaks chiral symmetry, can be of the same order of magnitude as the leading contribution, cf.\ Fig.~\ref{fig_disc}. The ``jumping'' of the points nicely demonstrates the strong dependence of the lattice artifacts on the chosen direction. In particular, the points along the axes [e.g., $(1,0,0)$], corresponding to the crosses in Fig.~\ref{fig_disc}, exhibit the largest discretization effects, while the points along the diagonal [e.g., $(1,1,1)$] are much better behaved. This fits in with earlier observations for correlation functions~\cite{Gimenez:2004me,Cichy:2012is,Tomii:2016xiv} and quark propagators in momentum space~\cite{Oliveira:2018lln}. The large contribution of the chiral odd part of the propagator is a peculiarity of using Wilson fermions, while large discretization effects at short distances are probably a general feature of all position space methods.\par%
The appearance of large contributions from the chiral odd part of the propagator would lead to huge lattice artifacts in the correlator. Therefore, we construct linear combinations of the correlation functions defined in Eqs.~\eqref{decomposition} where the chiral odd part of the propagator drops out to leading order in perturbation theory:%
\begin{align}%
&\frac12\bigl(T_{\rm SP} + T_{\rm PS}\bigr) \,, && \frac12\bigl(T_{\rm VA} + T_{\rm AV}\bigr) \,, && \frac12\bigl(T_{\rm VV} + T_{\rm AA}\bigr) \,.
\end{align}%
For the scalar-pseudoscalar correlator with a pion, this is equivalent to taking the real part (cf.\ Ref.~\cite{Bali:2017gfr}).\par
The discretization effects in the chiral even part of the propagator (these correspond to the blue points in Fig.~\ref{fig_disc}) are addressed as follows: we discard data points where the free field discretization effect exceeds $10\%$. This cut mainly excludes very short distances ($|\mathbf z|\lesssim 2a$) and directions close to the lattice axes. For the remaining data points, we define a correction factor $c^\text{corr}(z)$ such that the corrected propagator%
\begin{align}
 G^{\text{corr}}_{\text{latt}}(z)&\equiv c^\text{corr}(z) G_{\text{latt}}(z) \,,
\end{align}%
satisfies the condition%
\begin{align}
 \operatorname{tr}\bigl\{\slashed z G^{\text{corr}}_{\text{latt}}(z)\bigr\} &\overset{!}{=} \operatorname{tr}\bigl\{\slashed z G_{\text{cont}}(z) \bigr\} \,, \label{improvement_condition}
\end{align}%
where the trace runs over spin and color indices. To zeroth order accuracy in $\alpha_s$ (where $G^{\phantom{\mathrlap{\text{free}}}}_{\text{latt}}=G^{\text{free}}_{\text{latt}}$ is the free propagator), this leads to%
\begin{align}
 c^{\text{corr}}(z) &= \Biggl(\!\operatorname{tr}_D\bigl\{\slashed z G^{\text{free}}_{\text{latt}}(z)\bigr\} \frac{z^2 \pi^2}{2}\!\Biggr)^{-1} \!\! \frac{(-m^2 z^2)}{2} K_2\Bigl(m\textstyle\sqrt{-z^2}\Bigr) \,.
\end{align}%
This corresponds to multiplying the blue data points of Fig.~\ref{fig_disc} by factors so that in the noninteracting case the continuum result is retrieved. One should note that this procedure can only tame distance-dependent discretization effects. However, there are also momentum-dependent discretization effects, which are not taken into account. It is therefore no surprise that we still find particularly large discretization effects for the high momentum data at small distances. We have therefore decided to include only data points with $|\mathbf z|\geq 3a\approx\unit{0.21}{\femto\meter}$, which, setting the scale to $\mu=2/|\mathbf z|$, corresponds to $\mu\lesssim\unit{1.84}{\giga\electronvolt}$.\par
Finally, we remark that the pseudoscalar and scalar currents are (up to small mass-dependent effects) automatically order $a$ improved. In principle, we could also have order $a$ improved the axialvector and the vector currents. However, the improvement of $\gamma_\mu\gamma_5$ and of $\gamma_\mu$ would have required us to compute three-point functions with two currents situated at nonequal times (as well as a tensor current in the latter case).
\section{Results}\label{sect_results}%
\subsection{Parameter choices and first data survey}%
\begin{table}[tb]%
\centering%
\caption{\label{tab_momenta} The lattice momenta used in the analysis, where $\mathbf p=\frac{2\pi}{L}\mathbf n_{\mathbf p}$. $N_{\text{conf}}$ is the number of analyzed configurations, and $N_{\text{bin}}$ is the bin size used to reduce autocorrelations. Note that for the smallest momentum we have used only every tenth configuration.}
\begin{ruledtabular}
\begin{tabular}{c c c c}
$\mathbf n_{\mathbf p}$ & $|\mathbf p|$ & $N_{\text{conf}}$ & $N_{\text{bin}}$ \\ \hline
$\pm(\phantom{-}1,\phantom{-}0,\phantom{-}0)$ & $\unit{0.54}{\giga\electronvolt}$ & $\phantom{0}200$ & $\phantom{0}2$ \\
$\pm(\phantom{-}2,\phantom{-}0,\phantom{-}0)$ & $\unit{1.08}{\giga\electronvolt}$ &           $2000$ &           $20$ \\
$\pm(\phantom{-}2,\phantom{-}2,\phantom{-}0)$ & $\unit{1.53}{\giga\electronvolt}$ &           $2000$ &           $20$ \\
$\pm(\phantom{-}2,\phantom{-}2,\phantom{-}2)$ & $\unit{1.88}{\giga\electronvolt}$ &           $2000$ &           $20$ \\
$\pm(\phantom{-}3,\phantom{-}2,\phantom{-}1)$ & $\unit{2.03}{\giga\electronvolt}$ &           $2000$ &           $20$ \\
$\pm(\phantom{-}2,          -1,\phantom{-}3)$ & $\unit{2.03}{\giga\electronvolt}$ &           $2000$ &           $20$
\end{tabular}
\end{ruledtabular}
\end{table}%
\begin{figure}[tbp]
\includegraphics[width=\columnwidth]{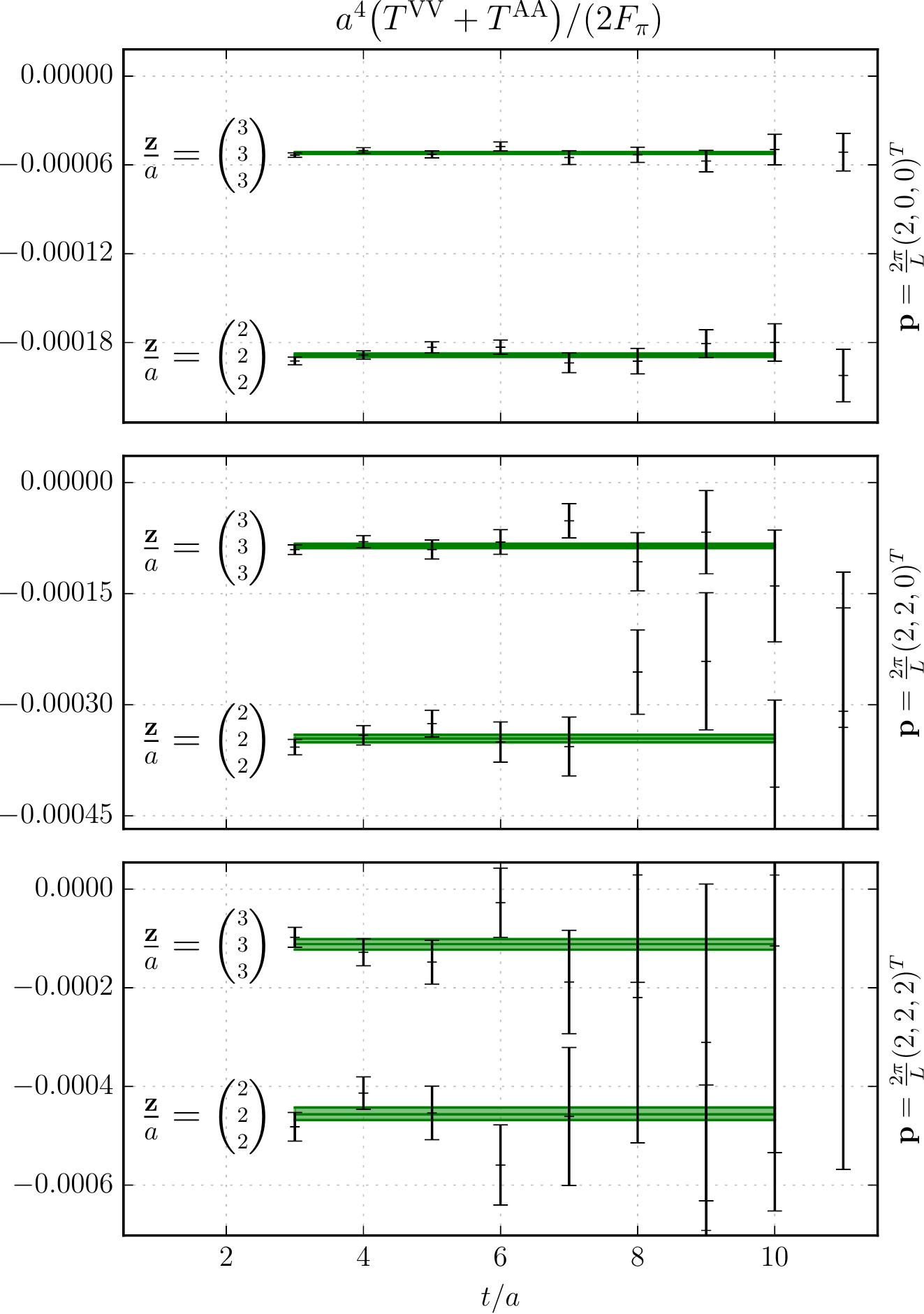}
\caption{\label{plateaus}The ratio~\eqref{eq:rato} for the example of the $\rm VV+AA$ combination of currents for different distances and momenta, together with our fitted results.}
\end{figure}
\begin{figure*}[tbp]%
\includegraphics[width=\columnwidth]{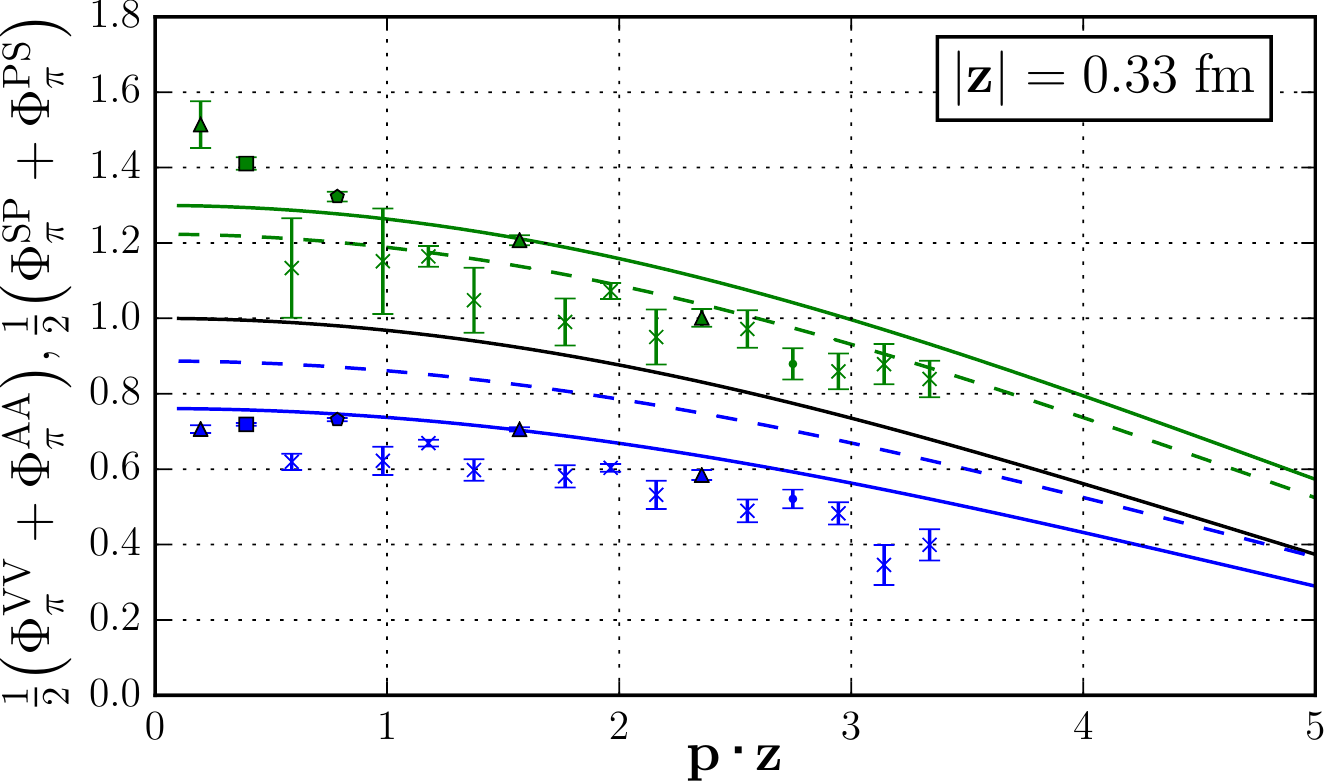}\hfill%
\includegraphics[width=\columnwidth]{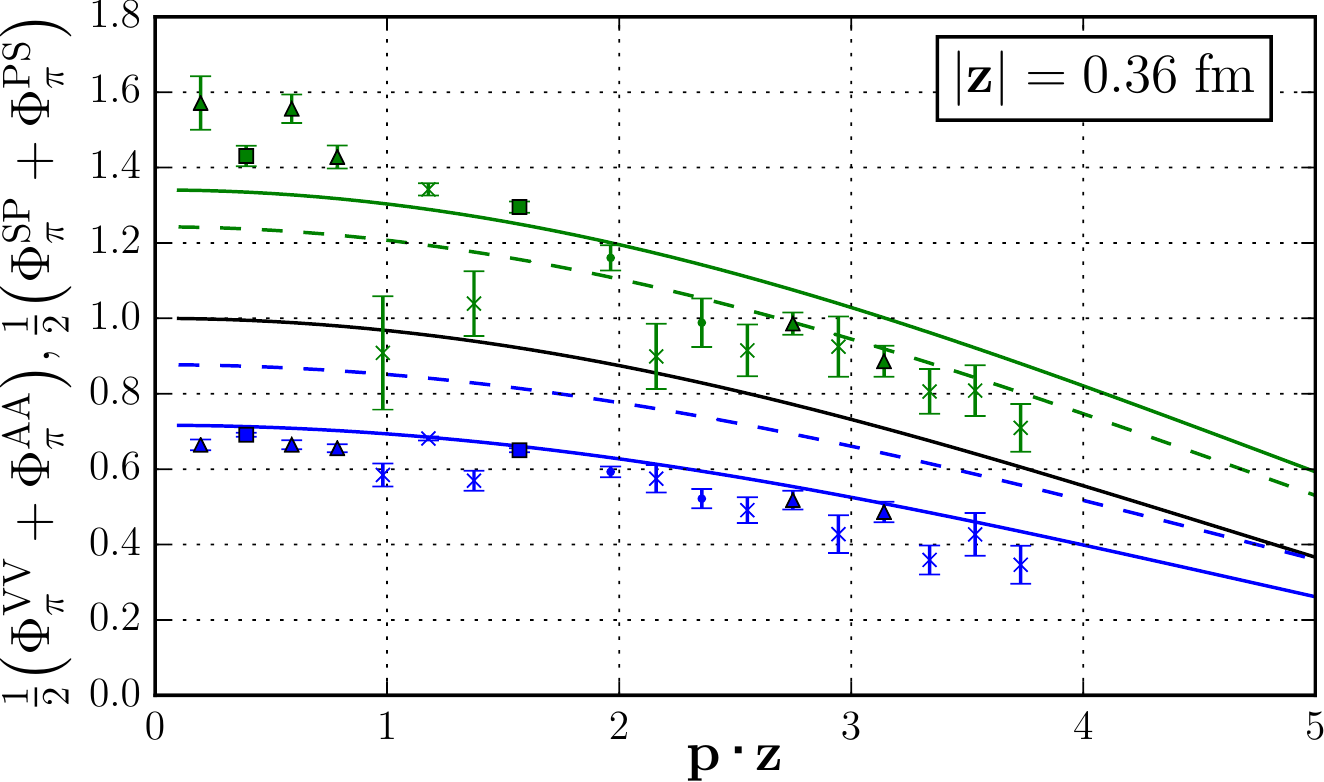}\\[0.085cm]%
\caption{\label{fig_HT}%
Data for the position space DA at two distances compared with expectations obtained using the second Gegenbauer coefficient $a_2^\pi(\unit{2}{\giga\electronvolt})=0.1364$ determined in Ref.~\cite{Braun:2015axa} with the moment method. The central solid curve corresponds to the (channel-independent) tree-level result at leading twist. The dashed lines include one-loop perturbative corrections for the two channels, and the outer solid lines also include higher-twist contributions [obtained using the QCD sum rule estimate $\delta^\pi_2(\unit{2}{\giga\electronvolt})=\unit{0.17}{\giga\electronvolt\squared}$ for the higher-twist normalization constant]. The upper data (green) are $\rm SP+PS$, and the lower data (blue) are $\rm VV+AA$.}
\end{figure*}%
Our analysis includes six different pion momenta ($12$, if one counts $\pm \mathbf p$ separately) with absolute values up to $|\mathbf p|=\unit{2.03}{\giga\electronvolt}$, cf.~Table~\ref{tab_momenta}. For the largest momentum, we have analyzed two different directions to increase statistics. Reaching such a large hadron momentum is quite challenging and was achieved by the combination of the momentum smearing technique, which enhances the overlap of the interpolating current with hadrons at large momenta, and the use of stochastic estimators described in Sec.~\ref{subsect_stochastic_estimation}, which allows us to take a volume average at the cost of additional stochastic noise. The latter trade-off turns out to be very advantageous and yields a significant reduction of the statistical errors compared to the sequential source method used in Ref.~\cite{Bali:2017gfr}.\par%
\begin{figure}[p]%
\includegraphics[width=\columnwidth]{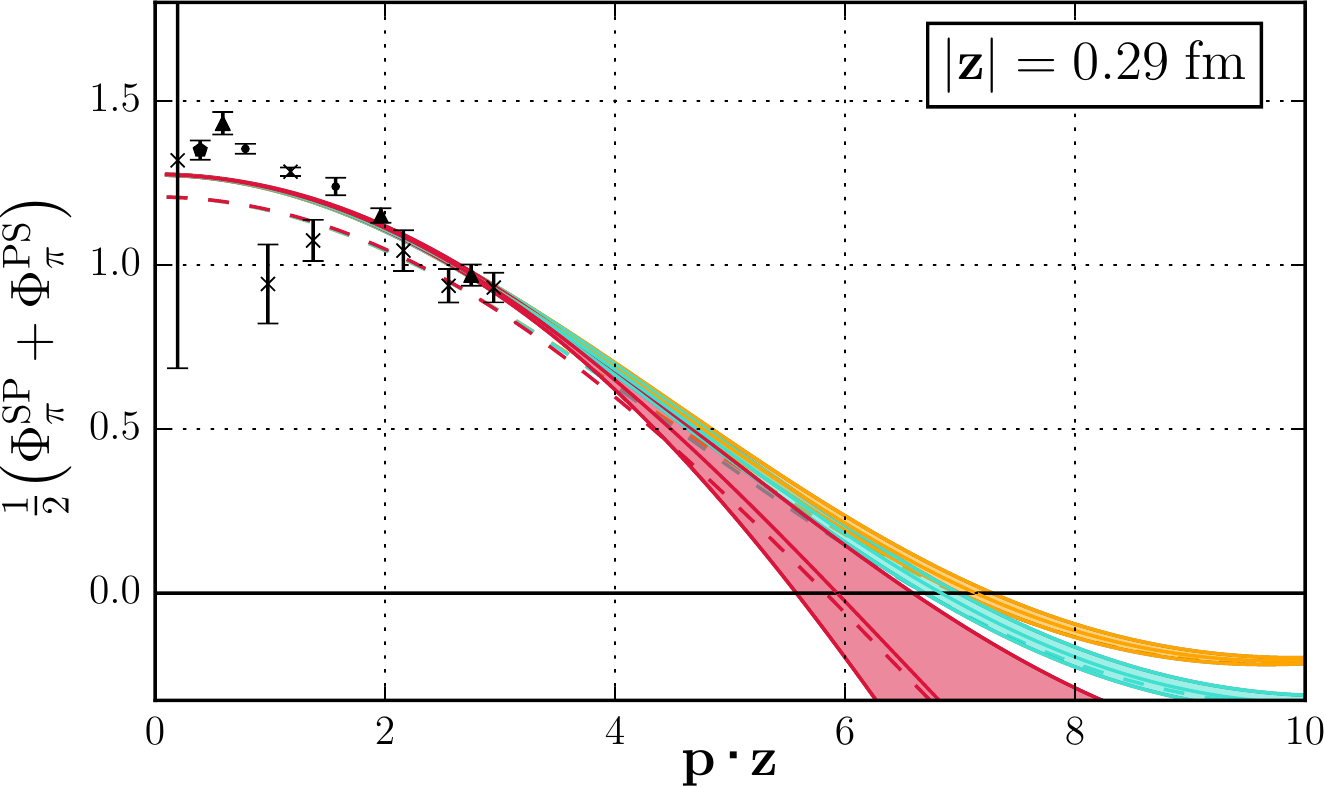}\\[0.085cm]
\includegraphics[width=\columnwidth]{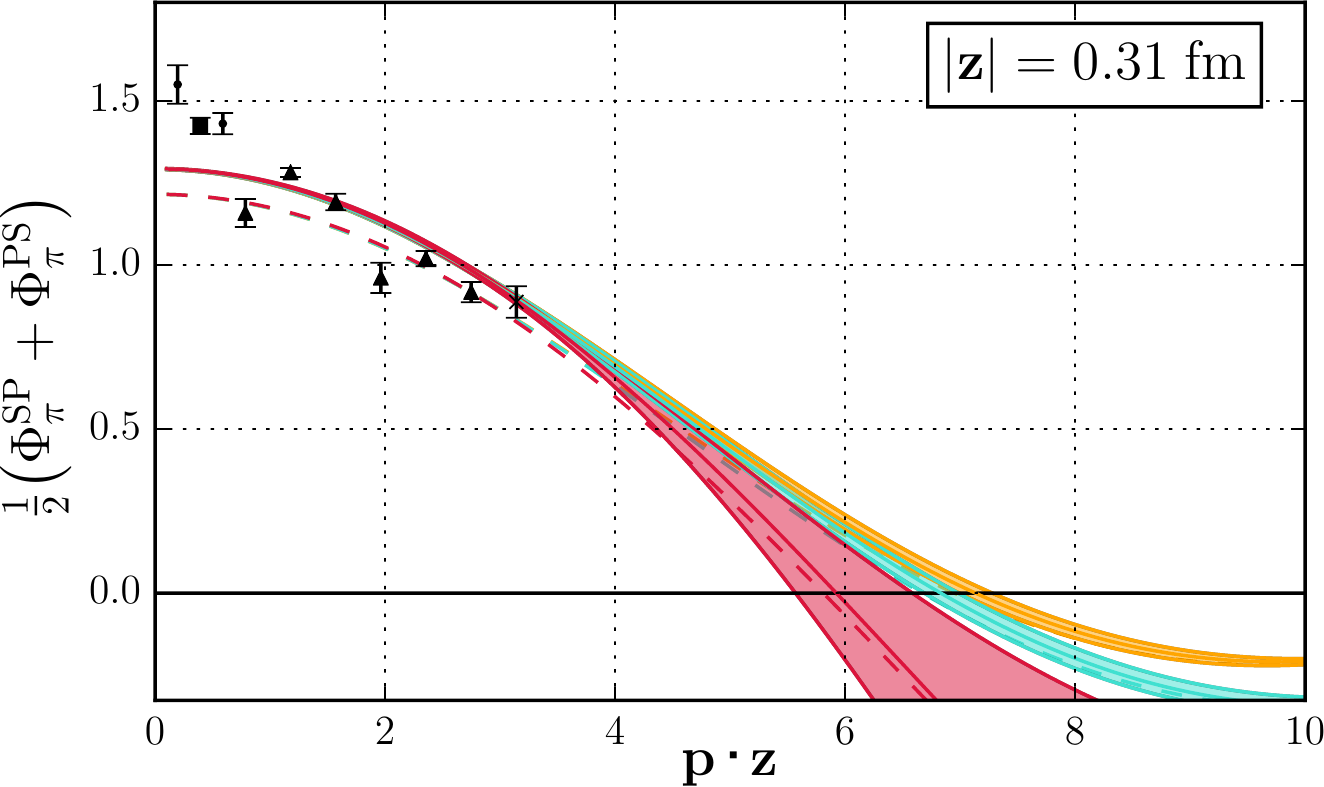}\\[0.085cm]
\includegraphics[width=\columnwidth]{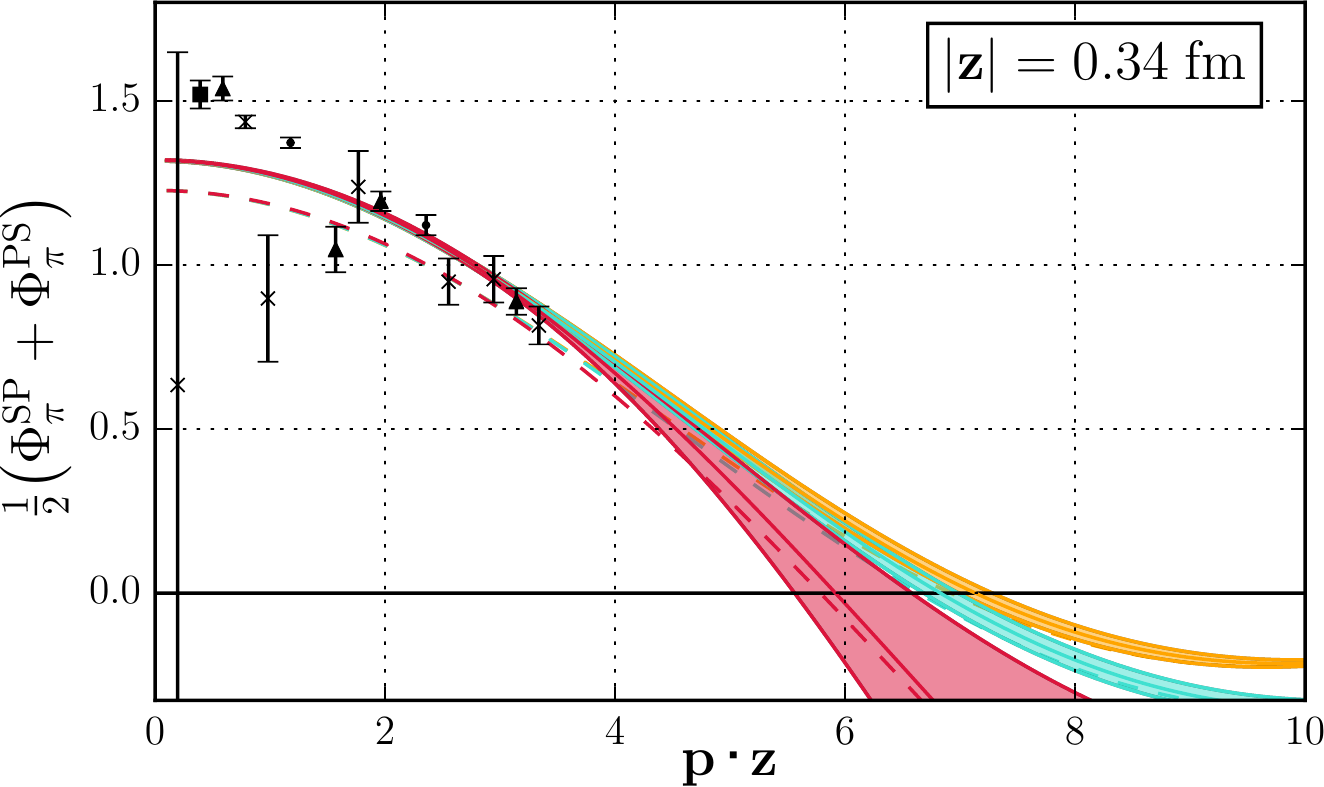}\\[0.085cm]
\includegraphics[width=\columnwidth]{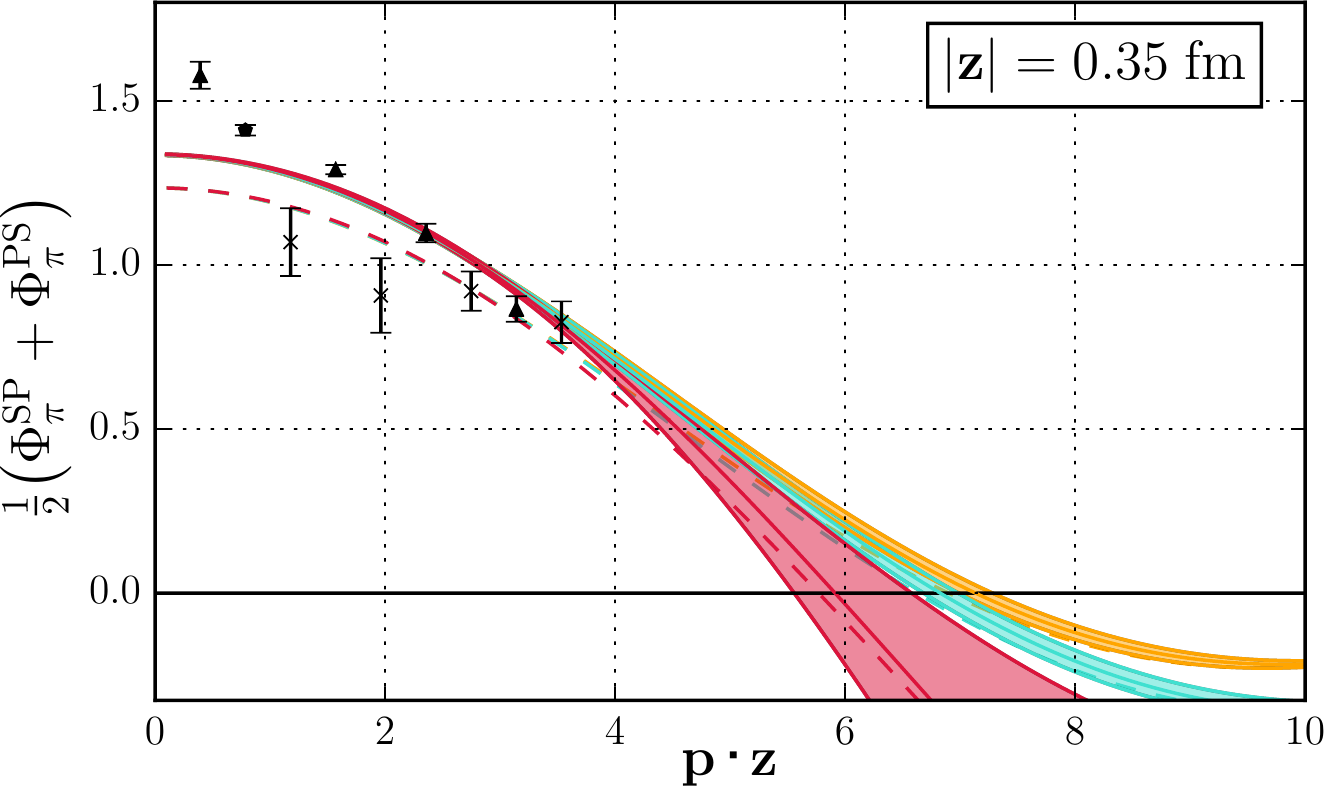}%
\caption{\label{fig_SP}%
The $\rm SP+PS$ correlator as a function of $\mathbf{p}\cdot \mathbf{z}$ for four different separations between the currents. The orange, turquoise, and red bands correspond to fits using the parametrizations A, B, and C explained in the text, cf.~Table~\ref{tab_results}. The dashed lines are obtained by subtracting the higher-twist contributions from the parametrizations.}
\end{figure}
\begin{figure}[p]%
\includegraphics[width=\columnwidth]{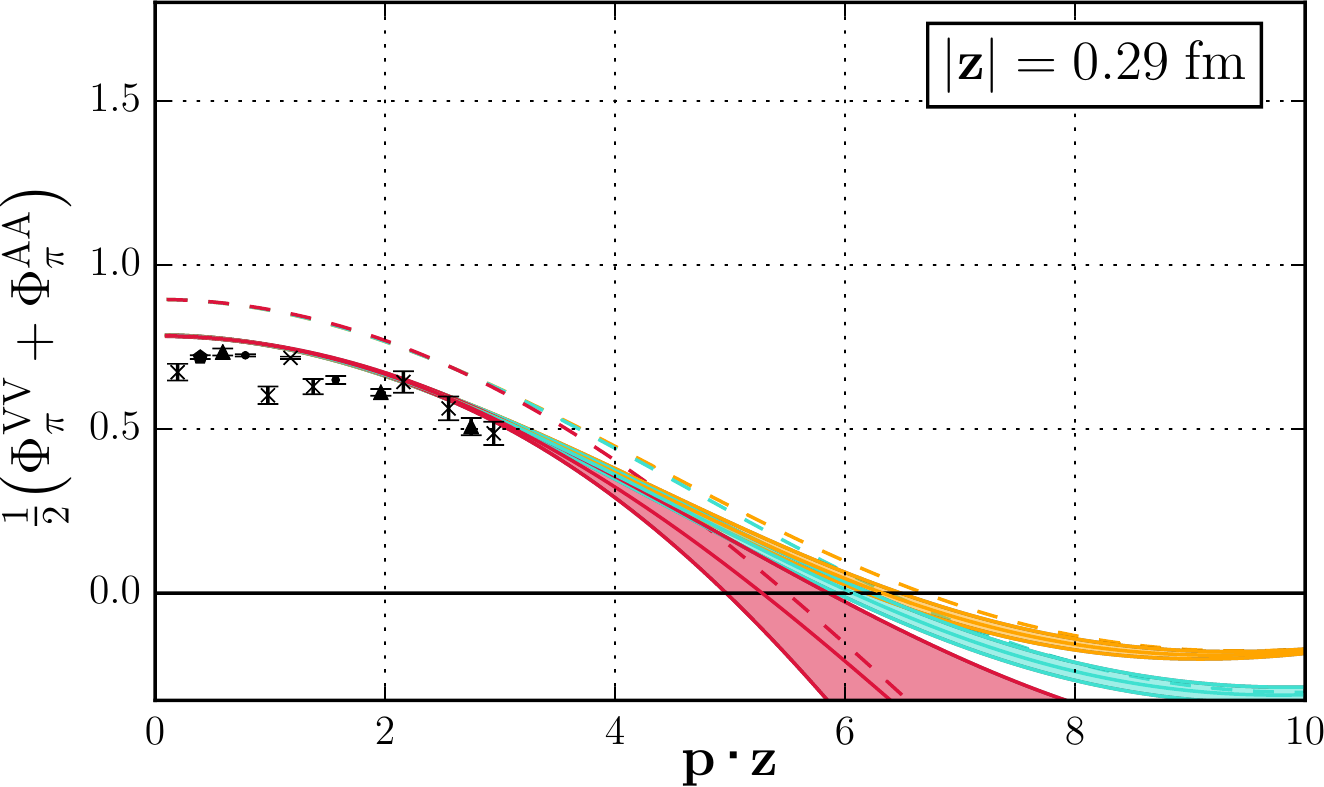}\\[0.085cm]
\includegraphics[width=\columnwidth]{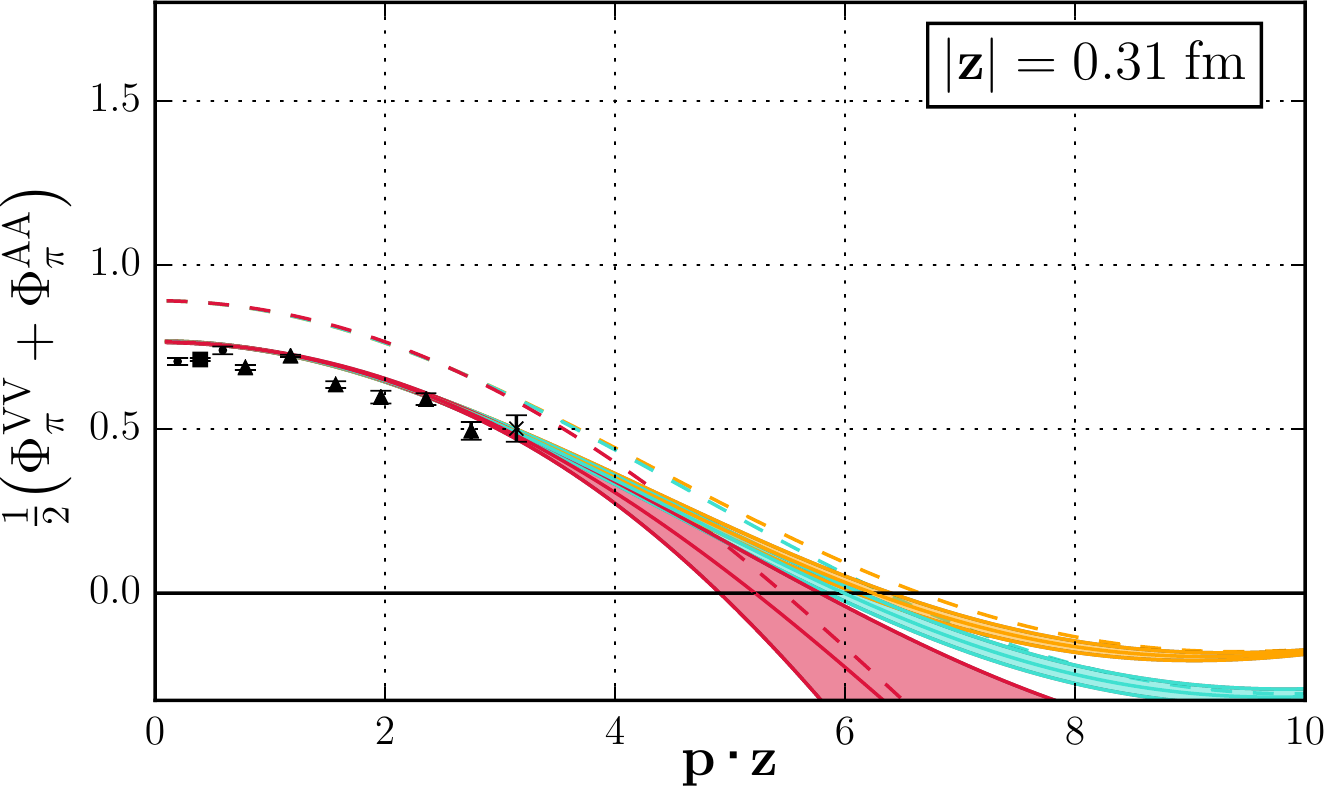}\\[0.085cm]
\includegraphics[width=\columnwidth]{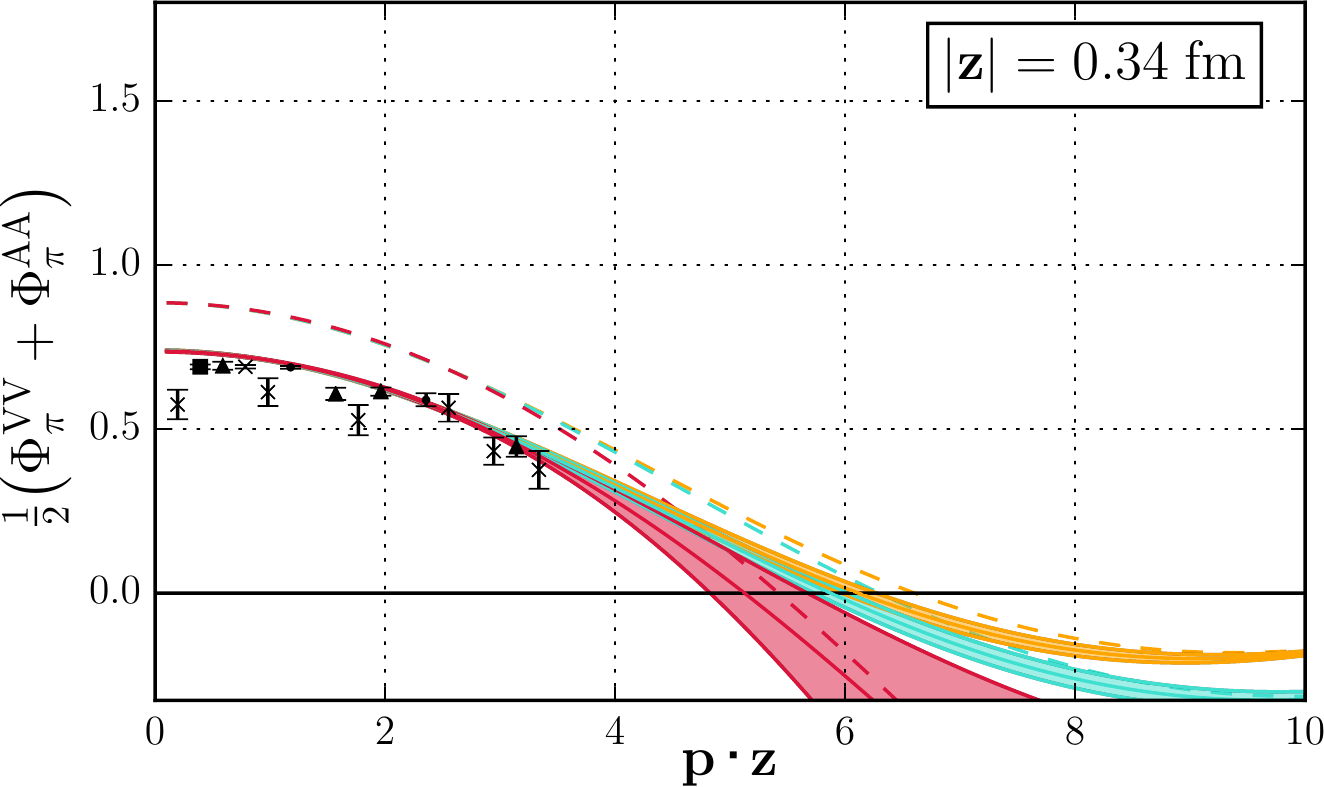}\\[0.085cm]
\includegraphics[width=\columnwidth]{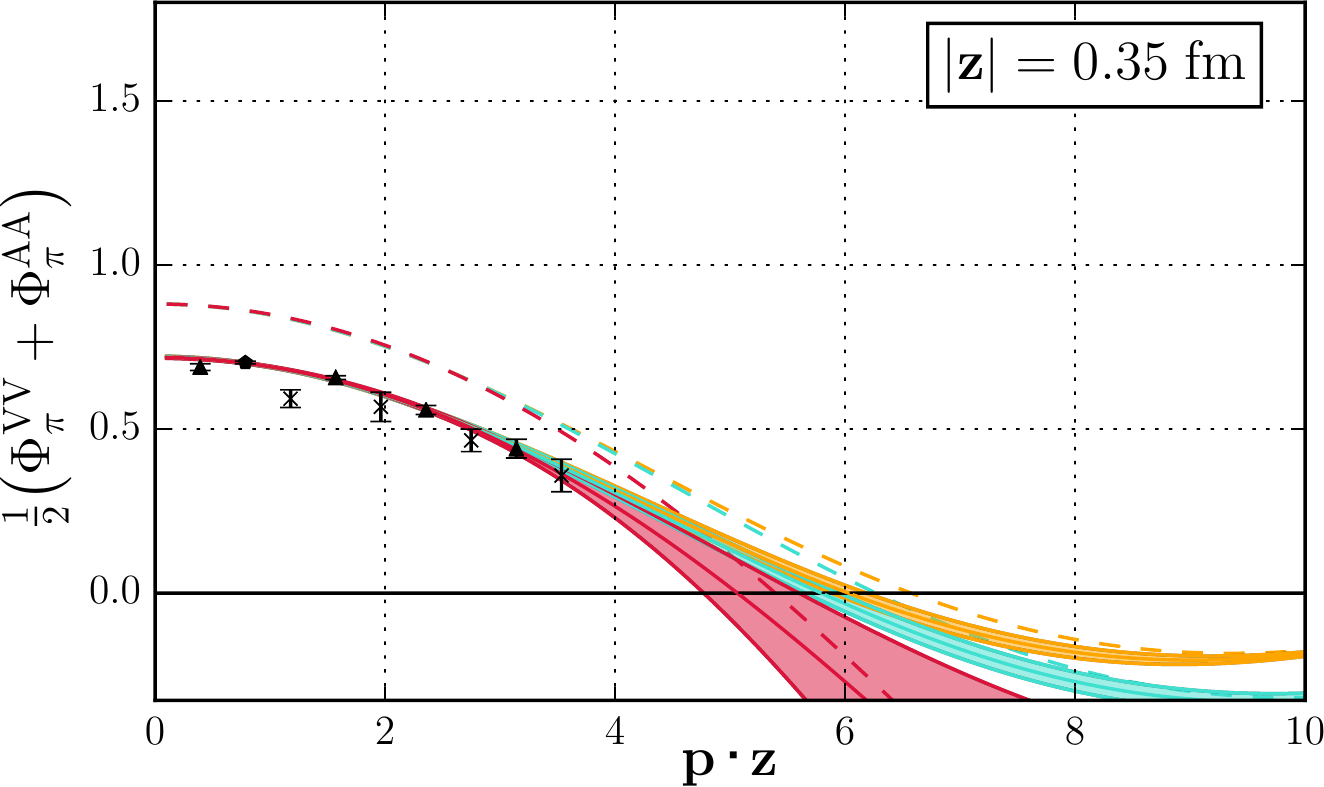}%
\caption{\label{fig_VV}%
The same as in Fig.~\ref{fig_SP}, but for the $\rm VV+AA$ correlation function.}
\end{figure}
\begin{figure}[p]%
\includegraphics[width=\columnwidth]{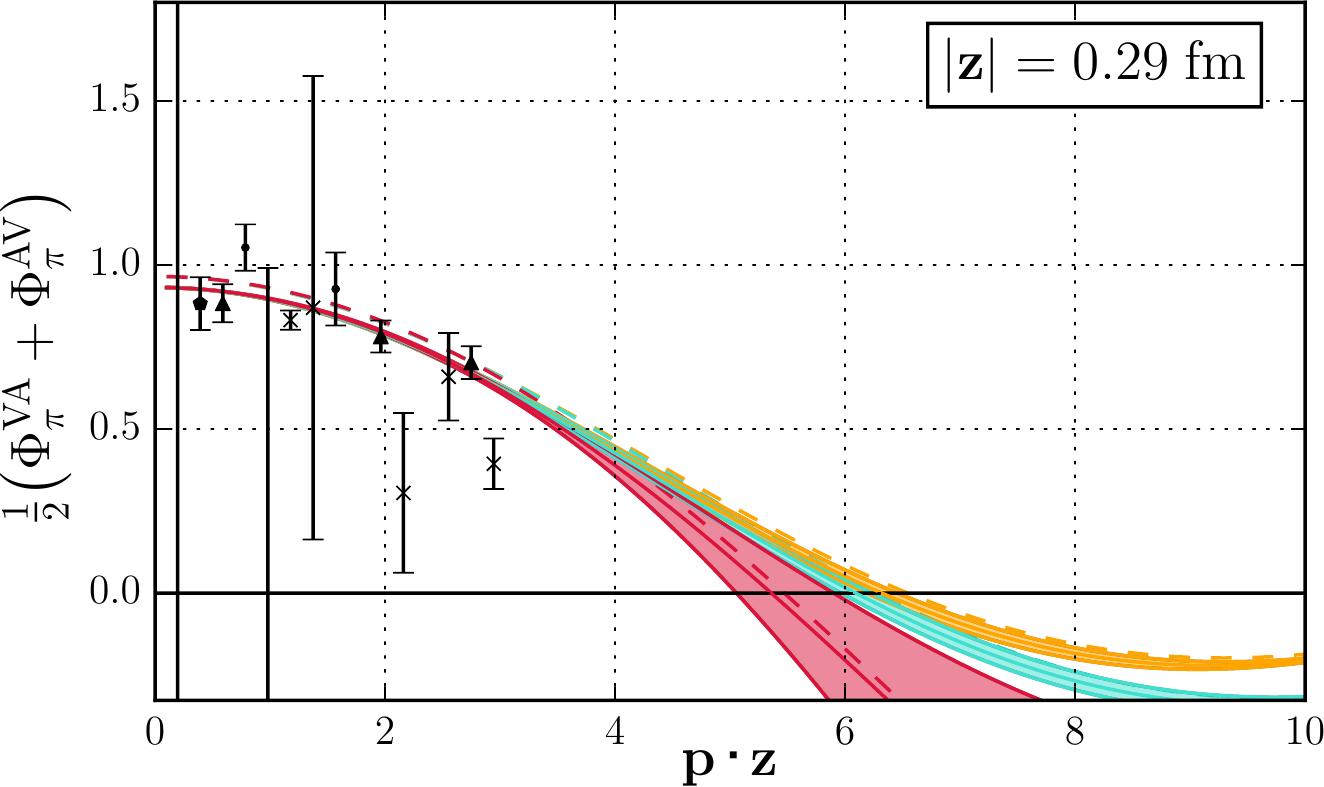}\\[0.085cm]
\includegraphics[width=\columnwidth]{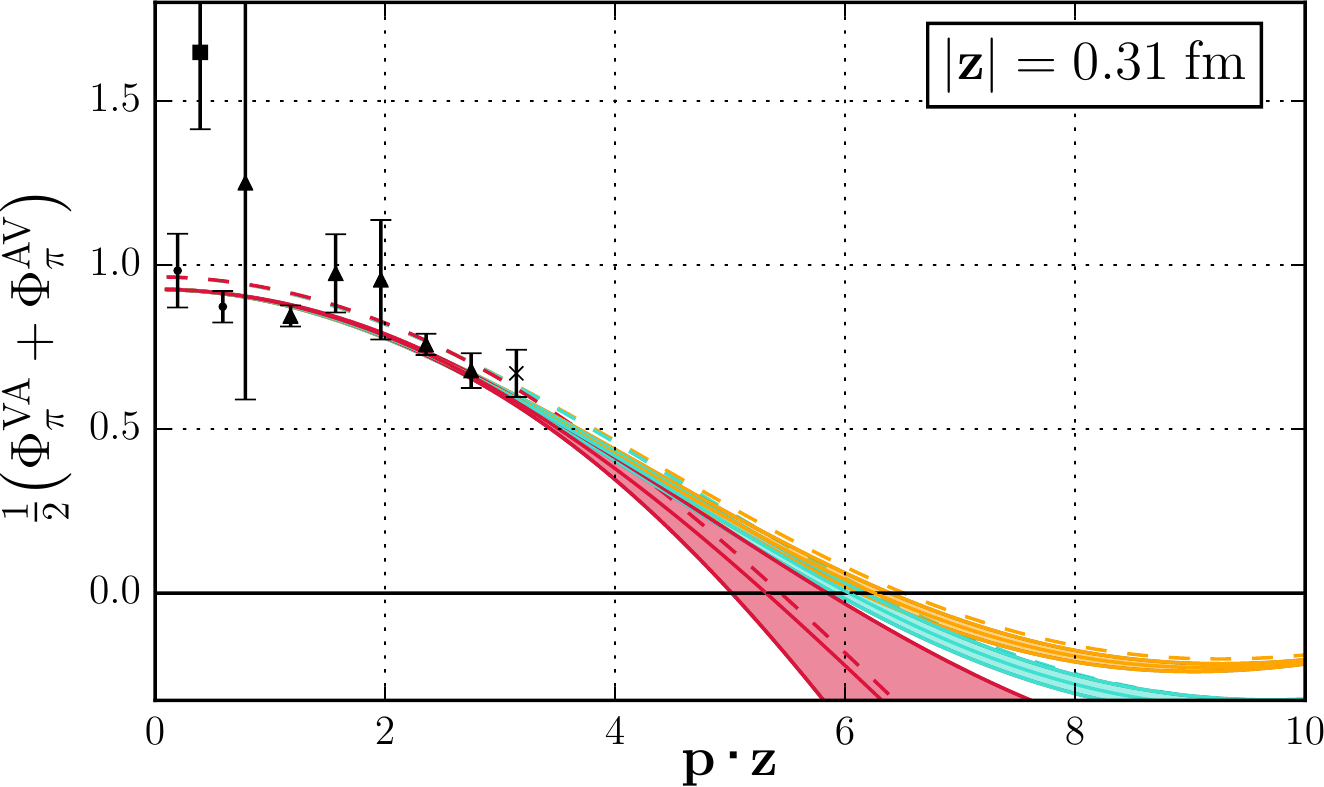}\\[0.085cm]
\includegraphics[width=\columnwidth]{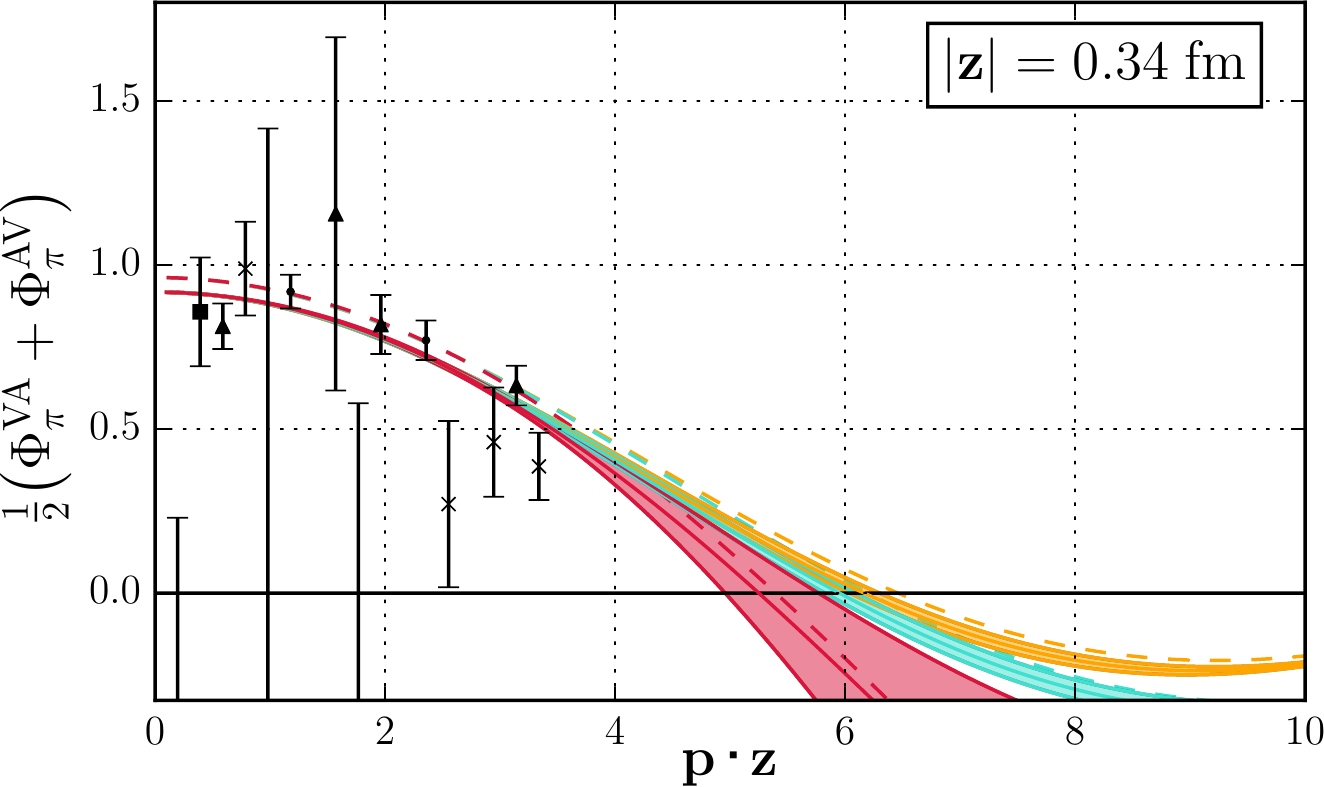}\\[0.085cm]
\includegraphics[width=\columnwidth]{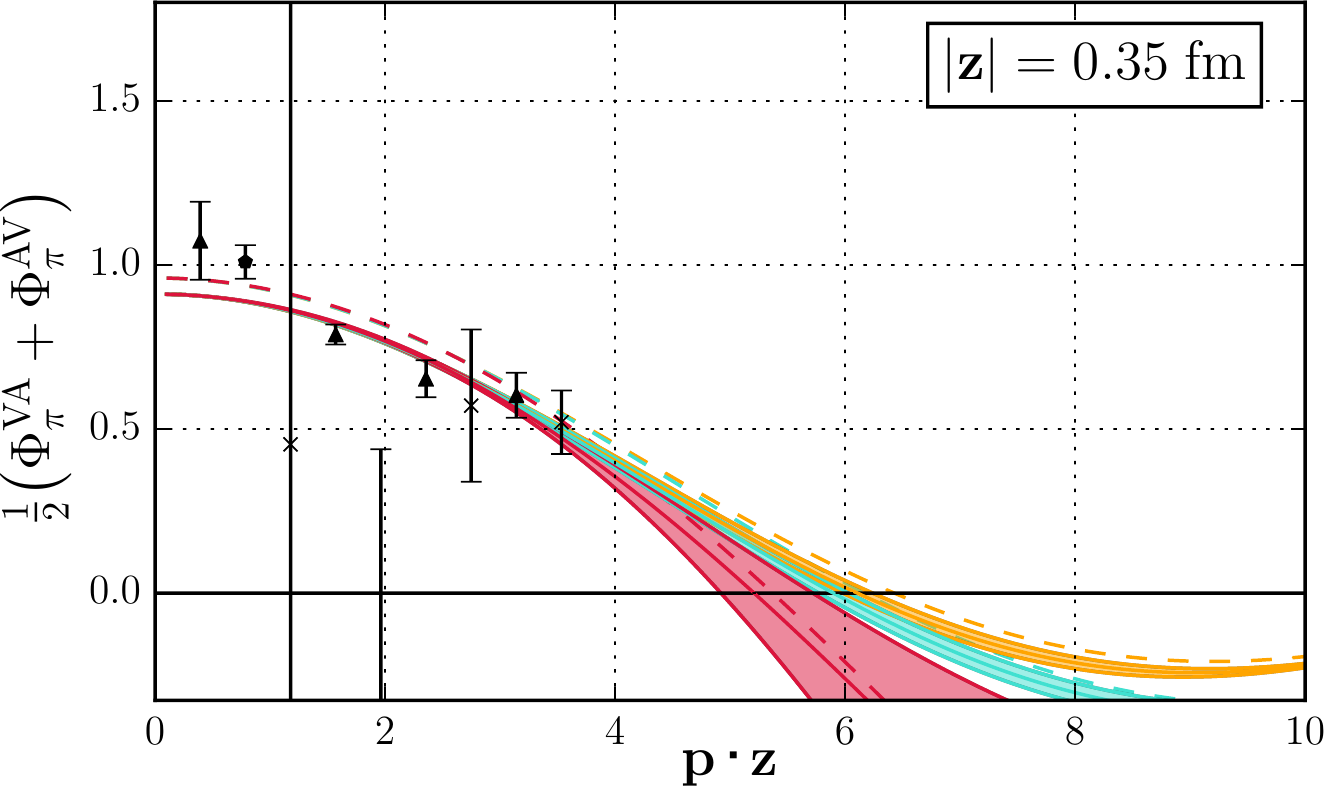}%
\caption{\label{fig_VA}%
The same as in Fig.~\ref{fig_SP}, but for the $\rm VA+AV$ correlation function.}
\end{figure}
Since the lattice data are analyzed using QCD factorization in the continuum, we are bound to using sufficiently small separations between the currents to ensure that the coefficient functions are perturbatively calculable. Together with the requirement of controllable discretization effects (see the previous section), this leaves us with the relatively narrow range of possible distances $ \unit{0.21}{\femto\meter} \lesssim|\mathbf{z}| \lesssim \unit{0.39}{\femto\meter}$, or $3a \leq |\mathbf z| \leq 5.5 a$ in units of the lattice spacing $a \simeq \unit{0.07}{\femto\meter}$. Since the direction of $\mathbf{z}$ is arbitrary, this constraint still allows for a large data set with ten different values for $|\mathbf{z}|$.\par
First, however, we should check if the ratios of three-point over two-point functions~\eqref{eq:rato} approach their asymptotic limits. We demonstrate this for the combination $(T_{\rm VV} + T_{\rm AA})/2$ for different momenta and distances in Fig.~\ref{plateaus}. Clearly, the momentum smearing was extremely successful in removing excited state contributions. Moreover, these seem to affect the two-point function in a similar way as the three-point functions, enabling additional cancellations to take place. The other channels exhibit a very similar behavior so that we can confidently fit to extended plateaus.\par
Next, in Fig.~\ref{fig_HT}, we compare our results at two typical distances for two different channels with the expectation obtained using the second Gegenbauer coefficient $a_2^\pi(\unit{2}{\giga\electronvolt})=0.1364$ that has been determined in Ref.~\cite{Braun:2015axa} with the moment method. The leading-twist position space DA (central solid line) is universal for all channels. The dashed lines include our one-loop perturbative corrections, while the solid lines also include higher-twist effects using the QCD sum rule estimate $\delta^\pi_2(\unit{2}{\giga\electronvolt})=\unit{0.17}{\giga\electronvolt\squared}$~\cite{Novikov:1983jt,Braun:1989iv,Ball:2006wn}. Unsurprisingly, toward the larger distance $|\mathbf{z}|$, both correction terms become more significant. The sign and magnitude of the predicted splitting are in good agreement with our data. However, there are quantitative differences: our data still show residual discretization effects, the models for the leading-twist and higher-twist DAs may not be correct, and there will be two-loop perturbative corrections as well. For the distances shown, the corrections to the leading order leading-twist DA are about 25\% in size, while even at $|\mathbf{p}\cdot\mathbf{z}|=4$, the differences between the models plotted in Fig.~\ref{fig:pionDAmodels} only amount to about 10\%; i.e., within our range of $z^2$ and $p\cdot z$ values, we are more sensitive to higher-twist effects and perturbative corrections than we are to the shape of the leading-twist DA. This is also expected from Fig.~\ref{fig-CPW} and the discussion of Sec.~\ref{sect_discussion}.\par
The data points in Fig.~\ref{fig_HT} as well as in Figs.~\ref{fig_SP}--\ref{fig_VA} below are obtained by performing a weighted average over all possible combinations of the distance $\mathbf z = (z_1,z_2,z_3)$ and momentum $\mathbf{p} = (p_1,p_2,p_3)$ that give the same values for the scalar product $|\mathbf p \cdot \mathbf z|$ and the same $\mathbf z^2$. The markers indicate how many different momenta from Table~\ref{tab_momenta} contribute to the average: $\text{dot}\mathrel{\hat=}1$, $\text{cross}\mathrel{\hat=}2$, $\text{triangle}\mathrel{\hat=}3$, $\text{square}\mathrel{\hat=}4$, $\text{pentagon}\mathrel{\hat=}5$, $\text{hexagon}\mathrel{\hat=}6$. The $\rm VV+AA$ channel yields the best signal by far, since in this case only one invariant structure that is consistent with the symmetries exists, and one can make use of an additional average over the open Lorentz indices, cf.\ Appendix~\ref{sect_QCDfactor_pro}. This averaging is not possible in the $\rm VA+AV$ channel since in this case one needs to project onto the specific leading-twist Lorentz structure Eq.~\eqref{projector_VA}. This projection entails a strong dependence of the signal-to-noise ratio on the momentum direction, and for some data points, none of the analyzed momenta yields a good signal. This explains the outliers with extremely large statistical errors in the figures below. Finally, the $\rm SP+PS$ channel, albeit slightly inferior to $\rm VV+AA$, also gives small statistical errors.\par
\subsection{Extraction of distribution amplitude parameters}
\begin{table}[tb]%
\centering%
\caption{\label{tab_results} Fit results for the Gegenbauer coefficients~$a_2^\pi$ and~$a_4^\pi$ as well as the higher-twist normalization constant~$\delta_2^\pi$. We consider three different DA parametrizations (which are all defined at the reference scale $\unit{2}{\giga\electronvolt}$) and various fit ranges in $\mu=2/|\mathbf{z}|$. Ansatz A corresponds to assuming the shape~\eqref{DA_alpha_parametrization}, while B and C use the expansion of the DAs in terms of Gegenbauer polynomials,~Eq.~\eqref{expansion_Gegenbauer}, truncated at $n=2$ and $n=4$. The numbers in parentheses give the statistical error. As discussed in the main text, a rather generous systematic uncertainty of 30\%--50\% should be assigned to these results and the values for $a_4^\pi$ from Ansatz A and B are meaningless. The fit range corresponding to the curves plotted in Figs.~\ref{fig_SP}--\ref{fig_VA} is highlighted.}
\begin{ruledtabular}
\begin{tabular}{lcE{1.5}E{2.6}E{1.6}l}%
&Ansatz	& \multicolumn{1}{c}{$a_2^\pi$} & \multicolumn{1}{c}{$a_4^\pi$} & \multicolumn{1}{c}{$\delta_2^\pi [\giga\electronvolt\squared]$} &  \\ \hline
\multicolumn{6}{c}{$\unit{0.9}{\giga\electronvolt}<\mu<\unit{1.8}{\giga\electronvolt}$} \\
&A		& 0.29(2) & \dcolcolor{gray} 0.16(2)  & 0.202(3) & $\alpha=0.17(5)$ \\
I&B		& 0.28(2) & \dcolcolor{gray} 0.0      & 0.202(3) & \\
&C		& 0.28(4) &                  0.0(0.6) & 0.202(4) & \\ \hline
\rowcolor{lightgray} \multicolumn{6}{@{}c@{}}{$\unit{1.0}{\giga\electronvolt}<\mu<\unit{1.8}{\giga\electronvolt}$} \\
&A		& 0.31(3) & \dcolcolor{gray} 0.17(2)  & 0.223(4) & $\alpha=0.13(5)$ \\
II&B		& 0.30(3) & \dcolcolor{gray} 0.0      & 0.223(4) & \\
&C		& 0.26(5) &                 -1.1(0.9) & 0.225(4) & \\ \hline
\multicolumn{6}{c}{$\unit{1.1}{\giga\electronvolt}<\mu<\unit{1.8}{\giga\electronvolt}$} \\
&A		& 0.36(3) & \dcolcolor{gray} 0.22(3)  & 0.242(4) & $\alpha=0.05(5)$ \\
III&B		& 0.35(3) & \dcolcolor{gray} 0.0      & 0.242(4) & \\
&C		& 0.29(6) &                 -1.6(1.2) & 0.244(4) & \\ \hline
\multicolumn{6}{c}{$\unit{1.0}{\giga\electronvolt}<\mu<\unit{1.5}{\giga\electronvolt}$} \\
&A		& 0.30(3) & \dcolcolor{gray} 0.17(2)  & 0.218(4) & $\alpha=0.15(5)$ \\
IV&B		& 0.30(3) & \dcolcolor{gray} 0.0      & 0.219(4) & \\
&C		& 0.22(5) &                 -1.7(0.9) & 0.222(4) & \\ \hline
\multicolumn{6}{c}{$\unit{1.0}{\giga\electronvolt}<\mu<\unit{1.3}{\giga\electronvolt}$} \\
&A		& 0.26(3) & \dcolcolor{gray} 0.14(2)  & 0.202(4) & $\alpha=0.22(6)$ \\
V&B		& 0.26(3) & \dcolcolor{gray} 0.0      & 0.202(4) & \\
&C		& 0.09(5) &                 -3.6(0.9) & 0.209(4) &
\end{tabular}%
\end{ruledtabular}%
\end{table}%
We are now in a position to analyze the whole data set and attempt to extract the pion DA, carrying out a global fit to all correlation functions using the expressions collected in  Sec.~\ref{sect_QCDfactor_res}. In Figs.~\ref{fig_SP}--\ref{fig_VA}, we show our data for four distances, along with such fits. The fits A (orange), B (turquoise), and C (red) correspond to different parametrizations of the leading-twist pion DA. Ansatz A corresponds to using the power-law parametrization~\eqref{DA_alpha_parametrization} with a free fit parameter $\alpha$, while B and C use the Gegenbauer expansion~\eqref{expansion_Gegenbauer} truncated at orders $n=2$ and $n=4$, respectively. All input parameters are taken at the reference scale $\mu_0=\unit{2}{\giga\electronvolt}$ and are evolved to $\mu = 2/|\mathbf{z}|$ using two-loop evolution equations, apart from the higher-twist parameter~$\delta^\pi_2$, where the scale dependence is taken into account at one-loop order. This means fits A and B have two free parameters --- $\alpha$, $\delta^\pi_2$~(A) and $a_2^\pi$, $\delta^\pi_2$~(B) ---  while fit C has three parameters: $a_2^\pi$, $a_4^\pi$, $\delta^\pi_2$. The results are shown in Table~\ref{tab_results} for different fit ranges in $2/|\mathbf{z}|$. The numbers in parentheses are the statistical errors, which turn out to be surprisingly small for $a_2^\pi$ and also for $\delta^\pi_2$. The Gegenbauer coefficient~$a_4^\pi$ cannot be constrained from our data, and including this contribution (Ansatz C compared to B) does not lead to a distinct improvement of the fit quality. The reason is obvious from Fig.~\ref{fig-CPW}, as the $n=4$ partial wave gives a negligible contribution to the correlation functions in the $\mathbf p \cdot \mathbf z$ range accessible in our study.\par%
The small statistical errors for $a_2^\pi$ and  $\delta^\pi_2$ are encouraging and allow us to analyze the (dominant) systematic errors. In order to gain some insight, we have performed the complete analysis for multiple fit ranges in the distance between the currents. A dependence on the lower bound in the distance (corresponding to larger scales) can indicate discretization effects, while a dependence on the upper bound shows the necessity to calculate higher-order corrections to the coefficient functions and, possibly, even higher-twist corrections. Such effects are clearly visible, cf.\ Table~\ref{tab_results}. As a second method to estimate the systematic uncertainty, one may assume that not-yet-calculated higher order perturbative effects are of the size of $\sim50\%$ of the one-loop correction. Both error estimation methods lead to the conclusion that, for the time being, one has to assign a systematic error of at least 30\%--50\% to the given numbers for $a_2^\pi$ and $\delta_2^\pi$, especially since other systematic uncertainties originating from an unphysically large pion mass as well as finite volume and lattice spacing corrections have not been addressed in this study.\par\clearpage%
\subsection{Discussion}%
Within the present range of distances and momenta, our data appear to be very sensitive to higher-twist corrections. These corrections can be quantified within our approach, and the corresponding parameter $\delta^\pi_2$ proves to be  only weakly correlated with the shape parameters of the pion DA. This can be explained as follows.\par%
First, it is crucial that perturbative and higher-twist corrections for the $\rm VV+AA$ and $\rm PS+SP$ correlators have similar magnitude and opposite sign, cf.~Fig.~\ref{fig_HT}. The higher-twist corrections contribute mostly to the difference of these two correlation functions, and much less to their sum. The effect of adding the $a_2^\pi$  parameter to the leading-twist pion DA is just the opposite; i.e., it affects both $\rm VV+AA$ and $\rm PS+SP$ correlators in a similar way. Second, writing the correlation functions $\Phi^{\rm XY}_\pi(p\cdot z,z^2)$ as an expansion in conformal partial waves similar to Eq.~(\ref{expansion_CPW}) for the DA, one can include higher-twist terms as contributions $\mathcal{O}(z^2)$ to the Gegenbauer coefficients; see Ref.~\cite{Braun:2007wv} for details. It turns out that this correction is largest for the leading term $a^\pi_0 \mapsto a^\pi_0(z^2) = a^\pi_0 + c \delta^\pi_2 z^2 + \ldots$ and affects $a^\pi_2$ and higher coefficients rather weakly. As a consequence, the higher-twist parameter $\delta^\pi_2$ can be extracted from position space correlators at small values of~$|\mathbf p\cdot \mathbf z|$, which explains its small statistical error.\par%
Note, however, that the obtained value is tied to using first order perturbative corrections $\mathcal{O}(\alpha_s)$ to the correlators, and will likely decrease if further terms are taken into account. This ambiguity is conceptual. It is related to the fact that matrix elements of twist-$4$ operators have quadratic power divergences already in the continuum theory and at the same time the perturbative series in leading twist in the minimal subtraction scheme suffers from factorial divergences (renormalons). One can show~\cite{Beneke:1998ui} that these two deficiencies are related and are cured in the sum of perturbative (leading-twist) and nonperturbative (higher-twist) effects. The higher-twist contribution, strictly speaking, should be viewed as an effective parametrization of the sum of the uncalculated higher orders of perturbation theory and ``genuine'' higher-twist effects; their separation requires additional regularization and is not necessary in the present context.\par%
Our result for $a_2^\pi$ has good statistical accuracy and all parametrizations of the DA lead to similar values that are somewhat larger than the result from the direct calculation of the second moment in Ref.~\cite{Braun:2015axa}, $a_2^\pi = 0.1364(154)(145)$ (at $\unit{2}{\giga\electronvolt}$). This should not be viewed as a contradiction as the systematic errors in the present study are not yet under control. They will decrease significantly in the future, especially if one could reach values of $|\mathbf p \cdot \mathbf z| \gtrsim 5$, which would also allow us to start probing the next Gegenbauer coefficient, $a_4^\pi$.\par%
\begin{figure}[tb]%
\includegraphics[width=\columnwidth]{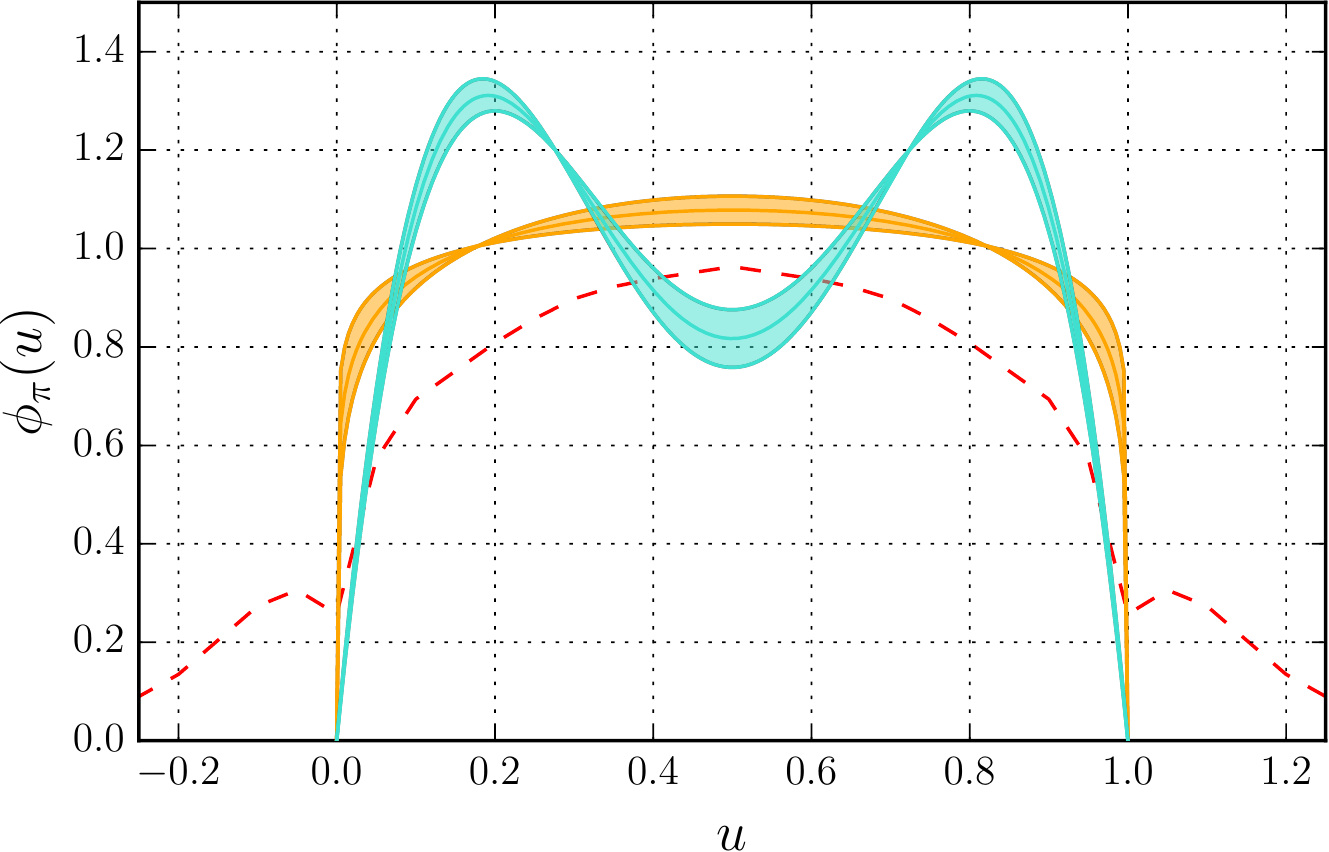}%
\caption{\label{fig_fitted_DA} The orange and turquoise bands correspond to DAs at a reference scale $\mu_0=\unit{2}{\giga\electronvolt}$ obtained from the fits to parametrizations A and B for fit range II, cf.\ Table~\ref{tab_results}. Both DAs lead to an equally good description of our data because they have a similar second Gegenbauer coefficient $a_2^\pi$, which is the only physically relevant information needed from the DA at the available range of $\mathbf p\cdot \mathbf z$. Note that the error only includes the statistical error for the used fit range and that the systematic uncertainty is considerably larger. For comparison, we have also included a result obtained using the quasidistribution approach (dashed line) taken from Ref.~\cite{Chen:2017gck}.}
\end{figure}%
The leading-twist DAs obtained from Ans\"atze~A and~B with fit range II are plotted in Fig.~\ref{fig_fitted_DA}. Note that the error bands only show the statistical error and that the systematic uncertainty (cf.\ fit range variation in Table~\ref{tab_results}) is considerably larger. Both DAs shown in Fig.~\ref{fig_fitted_DA} are in perfect agreement with our data since they yield similar values for $a_2^\pi$, which is, as discussed above, the only parameter that is relevant for the description of the data within the range of $\mathbf p\cdot \mathbf z$ that is currently available. In order to distinguish these DAs from each other, one would need data at larger $|\mathbf p\cdot \mathbf z|$ values that are sensitive to higher Gegenbauer coefficients. Our results favor DAs that, at a scale of $\unit{2}{\giga\electronvolt}$, are considerably broader than the asymptotic DA.\par%
\section{Conclusion and outlook}\label{sect_conclusion}%
In this work, we demonstrate that the method proposed in Ref.~\cite{Braun:2007wv} for the determination of collinear parton distributions does not only lead to qualitatively appealing results (see our first article on the topic~\cite{Bali:2017gfr}) but is indeed capable of producing quantitative results with surprisingly small statistical errors. The latter is possible due to the combination of momentum smearing (improving the signal for hadrons with large momentum) with stochastic estimation. A main characteristic of our approach is that we use an equal-time correlation function of two local currents, connected by a light quark propagator, instead of a nonlocal operator, connected by a Wilson line gauge transporter. This has multiple advantages:%
\begin{enumerate}
\item We circumvent problems originating from the renormalization of nonlocal operators entirely, since the local currents we use can be renormalized using well-tested standard methods.
\item Using a quark propagator easily allows us to evaluate distances that are not aligned with a lattice axis. While this is also possible when using a smeared Wilson line~\cite{Alexandrou:2018pbm}, the latter may interfere with the renormalization. On-axis separations are actually the worst-case scenario as we find discretization effects to be largest for these directions, cf.\ Fig.~\ref{fig_disc}. A restriction to the axes also implies a considerable reduction of the data set; in Figs.~\mbox{\ref{fig_SP}--\ref{fig_VA}}, this would correspond to having only one data point per momentum per plot.
\item We can evaluate multiple channels, which gives us an additional handle on the systematic error. It is crucial that higher-twist corrections for different correlation functions are related and can have opposite sign. The channels we have analyzed lead to consistent results and can be used in a global analysis to obtain values for the leading Gegenbauer coefficient of the leading-twist DA and for the higher-twist normalization constant. Note that it should be possible to include data from the Wilson line approach as an additional channel in such a global analysis.
\item Using two local currents instead of one nonlocal current has the nice feature that one can in principle apply the usual operator improvement within the Symanzik improvement program to remove $\mathcal O(a)$ effects.
\item For the matrix element with two local operators, finite volume effects have been calculated in Ref.~\cite{Briceno:2018lfj}. The results therein show that, even for an intermediate lattice size with $m_\pi L=4$ (in our case $m_\pi L\approx 3.4$), one has to expect large volume effects (\mbox{$\sim10\%$}) once the distance between the two currents approaches half of the lattice extent, i.e., if $|\mathbf z| \approx 0.5 L$. In this respect, it is helpful that our analysis method is restricted to relatively small distances $|\mathbf z|\leq 5.5a\approx 0.2 L$ where perturbative QCD is applicable, meaning that these volume effects are under control.
\end{enumerate}%
Another important feature of our analysis method is that we match the perturbative QCD calculation and the lattice data directly in position space. Note that such a position space analysis is not tied to using a light quark propagator but can also be performed within the Wilson line approach (see, e.g., Refs.~\cite{Radyushkin:2017cyf,Orginos:2017kos}). The obvious advantage over a quasidistribution-type analysis is that one can directly see on the data level (in Figs.~\ref{fig_SP}--\ref{fig_VA}) whether the perturbative matching between the off--light-cone correlation function one calculates and the light-cone quantities one is interested in actually works.\par%
From a global fit to our data, we obtain values for $a_2^\pi$ and $\delta_2^\pi$ with unexpectedly small statistical errors. An analysis of the fit range dependence showed that we have reached an accuracy where the systematic uncertainties by far dominate. Nevertheless, one can say that the value obtained for $a_2^\pi$ indicates a DA that, at $\unit{2}{\giga\electronvolt}$, is considerably broader than the asymptotic one. The value we obtain for the higher-twist matrix element~\eqref{def:delta-pi-2},%
\begin{equation}
\unit{0.2}{\giga\electronvolt\squared}\lesssim\delta_2^\pi\lesssim\unit{0.25}{\giga\electronvolt\squared}\,,
\end{equation}%
is only slightly larger than sum rule estimates, which lie at approximately $\delta_2^\pi=\unit{0.17}{\giga\electronvolt\squared}$ at a scale of $\unit{2}{\giga\electronvolt}$. To our knowledge, this is the first determination of $\delta_2^\pi$ from lattice QCD.\par%
We find that, restricting the analysis to distances where perturbation theory is applicable, even our largest momentum (with $|\mathbf p|=\unit{2.03}{\giga\electronvolt}$) is still slightly too small for the data to be sensitive to $a_4^\pi$. However, it is clear that this situation will improve dramatically if one could reach values of $|\mathbf p| > \unit{2.5}{\giga\electronvolt}$.\par%
Having reached small statistical errors only to find a large systematic uncertainty may seem a bit unsettling at first. In fact, the opposite is the case, since all main problems we have identified can be solved by systematically improving the analysis and provide us with some guidance toward the next necessary steps. On the lattice side of the calculation, we find discretization effects to be the gravest issue (despite all efforts to tame them described in Sec.~\ref{subsect_disc}). We plan to address this problem with a twofold strategy, both by drastically reducing the lattice spacing and, in the long run, by implementing $\mathcal O(a)$ improvement. To reduce the systematic uncertainty from the perturbative side of the calculation, our results clearly call for a two-loop analysis and a more systematic study of higher-twist effects.\par%
\begin{acknowledgments}%
This work has been supported by the Deut\-sche For\-schungs\-ge\-mein\-schaft (Grant No.\ SFB/TRR\nobreakdash-55), the Polish National Science Center (Grant No.\ \mbox{UMO-2016/21/B/ST2/01492}), and the Stu\-di\-en\-stif\-tung des deut\-schen Vol\-kes. We acknowledge PRACE (Grant No.\ 2016163989) for awarding us access to Marconi-KNL hosted by CINECA at Bologna, Italy. Part of the analysis was carried out on the QPACE~2~\cite{Arts:2015jia} Xeon~Phi installation of the SFB/TRR\nobreakdash-55 in Regensburg. We used a modified version of the {\sc Chroma}~\cite{Edwards:2004sx} software package along with the {\sc Lib\-Hadron\-Analysis} library and the multigrid solver implementation of Ref.~\cite{Heybrock:2015kpy} (see also Ref.~\cite{Frommer:2013fsa}).\par%
\end{acknowledgments}%
%
%%%%%%%%%%%%%%%%%%%%%%%%%%%%%%%%%%%%%%%%%%%%%%%%%%%%%%%%%%%%%%%%%%%%%%%%%%%%%
%%%%%%%%%%%%%%%%%%%%%%%%%%%%%   Appendix   %%%%%%%%%%%%%%%%%%%%%%%%%%%%%%%%%%
%%%%%%%%%%%%%%%%%%%%%%%%%%%%%%%%%%%%%%%%%%%%%%%%%%%%%%%%%%%%%%%%%%%%%%%%%%%%%
%
\appendix 
\section{Lorentz-projection operators}\label{sect_QCDfactor_pro}%
In order to project onto $T_{\rm XY}$ defined in Eq.~\eqref{decomposition}, we can use the projection matrices%
\begin{subequations}
\begin{align}
 P^{\rm VV}_{\mu\nu} &=\frac{\varepsilon_{\mu\nu\rho\sigma} p^\rho z^\sigma p\cdot z  }{2 i (\!  (p\cdot z)^2 - p^2 z^2 )} \, , \label{projector_VV} \displaybreak[0] \\
 P^{\rm VA}_{\mu\nu} &=\frac{p\cdot z}{2 (\! (p\cdot z)^2 - p^2 z^2 )^2} \bigl[ ( 2 (p\cdot z)^2 + p^2 z^2 ) ( p_\mu z_\nu + z_\mu p_\nu )  \nonumber  \\
 \MoveEqLeft[1]  - 3 p\cdot z ( z^2  p_\mu p_\nu +  p^2 z_\mu z_\nu)  - p\cdot z (\!  (p\cdot z)^2 - p^2 z^2 ) g_{\mu\nu}   \bigr] \, , \label{projector_VA} \\
 P^{\rm AA}_{\mu\nu} &=P^{\rm VV}_{\mu\nu} \, , \qquad P^{\rm AV}_{\mu\nu} =P^{\rm VA}_{\mu\nu} \,,
\end{align}%
\end{subequations}%
such that $T_{\rm XY}= P^{\rm XY}_{\mu\nu} \mathbb T^{\mu\nu}_{\rm XY}$. For the vector-axialvector channel this projection is the only possibility to obtain~$T_{\rm VA}$. In the case of the vector-vector channel (and the axialvector-axialvector channel), however, one can obtain $T_{\rm VV}$ (or $T_{\rm AA}$) from any channel with two fixed indices $\mu$ and $\nu$ as long as $\varepsilon^{\mu\nu\rho\sigma} p_\rho z_\sigma\neq0$:%
\begin{align}
 T_{\rm VV} &= \mathbb T^{\mu\nu}_{\rm V V}  \frac{-i p\cdot z}{\varepsilon^{\mu\nu\rho\sigma} p_\rho z_\sigma} \,.
\end{align}%
In our final analysis we use a weighted average of the individual channels, where the weight is defined as the inverse standard deviation squared of the respective channel. This yields a much better signal than the projection with~\eqref{projector_VV}, which basically averages over all vector-vector/axialvector-axialvector channels.\par%
\section{Higher-twist corrections}\label{App:HT}%
In this Appendix we provide some details on the calculation of the higher-twist corrections. First of all, for nonvanishing quark masses, also the chiral odd twist-$3$ pion DAs have to be taken into account:%
\begin{align}
\MoveEqLeft \langle 0 | \bar u(z) i\gamma_5 [z,-z] u(-z) | \pi^0(p)\rangle =
\nonumber\\ &=
\frac{F_\pi m_\pi^2}{2 m_u}\, \int_0^1 du \, e^{i(2u-1) p\cdot z}\,\phi^{p}_{3}(u) +\mathcal{O}(z^2)\,,
\nonumber\\
\MoveEqLeft \langle 0 | \bar u(z) \sigma_{\alpha\beta}\gamma_5 [z,-z] u(-z) |\pi^0 (p)\rangle =
\nonumber \\ 
& =-\frac{i}{3}\, \frac{F_\pi  m_\pi^2}{2 m_u}  (p_\alpha z_\beta-
p_\beta z_\alpha) \int_0^1 du \, e^{i(2u-1) p\cdot z}\,\phi^{\sigma}_{3}(u)
\nonumber\\
&\quad+\mathcal{O}(z^2)\,,
\end{align}%
They enter our calculation multiplied by the quark mass and become part of the pion mass correction. Since these contributions are small, we have used the simplest asymptotic expressions,%
\begin{align}
  \phi^{p}_{3}(u) &=1\,, & \phi^{\sigma}_{3}(u) &= 6u(1-u)\,,
\end{align}%
and omitted corrections due to the three-particle quark-antiquark-gluon DA~\cite{Braun:1989iv}. Complete expressions for the twist-$3$ matrix elements can be found in Refs.~\cite{Braun:1989iv,Ball:2006wn}.\par%
To twist-$4$ accuracy $\mathcal{O}(z^2)$ and omitting contributions of four-particle operators with two gluon fields and/or an extra quark-antiquark pair (which are expected to have very small matrix elements), one needs to consider two contributions shown schematically in Figs.~\ref{fig-pigammagamma}$(b)$ and \ref{fig-pigammagamma}$(c)$. The first of them is calculated using the background field expansion of the quark propagator~\cite{Balitsky:1987bk},%
\begin{widetext}%
\begin{align}
\label{propagator}
\contraction{}{q}{(z)}{\overline{q}}q(z) \overline{q}(-z) &=  \frac{i}{16\pi^2} \frac{\slashed{z}}{z^4}[z,-z]  
 -\frac{1}{32\pi^2z^2}
\int \limits_{-1}^1dv [z,vz]\bigg\{iz^\rho g\widetilde G_{\rho\sigma}(vz)\gamma^\sigma\gamma_5 + v  z^\eta gG_{\eta\rho}(vz)\gamma^\rho\biggr\}[vz,-z]  
\nonumber\\
&\quad- \frac{m_q}{16\pi^2}\frac{\mathds 1}{z^2} 
-\frac{i}4 m_q \int \frac{d^4k}{(2\pi)^4} \frac{e^{-2ik \cdot z}}{k^4} \int\limits_{-1}^1 dv\,[z,vz]gG^{\mu\nu}(vz)\sigma_{\mu\nu}[vz,-z] +\ldots\,,
\end{align}%
\end{widetext}%
where $[z_1,z_2]$ is the straight-line ordered Wilson line connecting the point $z_1$ to $z_2$, while $G$ ($\widetilde G$) denotes the (dual) field strength tensor. The last shown term is IR divergent and has to be regularized. It turns out, however, that this term does not contribute to our correlation functions and can be dropped.\par%
The necessary matrix elements can be parametrized in terms of the four higher-twist DAs~\cite{Braun:1989iv},%
\begin{align}
\MoveEqLeft \langle 0 | \bar u (z)\gamma_\mu\gamma_5
gG_{\alpha\beta}(vz)u (-z)|\pi^0(p)\rangle =
\nonumber\\
& = p_\mu (p_\alpha z_\beta - p_\beta z_\alpha)\, \frac{1}{p \cdot z}\, F_{\pi} \Phi_{4;\pi}(v,p \cdot z) 
\nonumber\\
&\quad + (p_\beta g_{\alpha\mu}^\perp - p_\alpha g_{\beta\mu}^\perp) F_{\pi} \Psi_{4;\pi}(v,p \cdot z) + \dots, \displaybreak[0]
\\
\MoveEqLeft \langle 0 | \bar u (z)\gamma_\mu i
g\widetilde{G}_{\alpha\beta}(vz)u(-z)| \pi^0(p)\rangle =
\nonumber\\
& = p_\mu (p_\alpha z_\beta - p_\beta z_\alpha)\, \frac{1}{p \cdot z}\, F_\pi
\widetilde\Phi_{4;\pi}(v,p \cdot z)
\nonumber\\
 &\quad + (p_\beta g_{\alpha\mu}^\perp -p_\alpha g_{\beta\mu}^\perp) F_{\pi} \widetilde\Psi_{4;\pi}(v,p \cdot z) + \dots,
\end{align}%
with the short-hand notation%
\begin{align}
{\cal F}(v,p \cdot z) = \int{\cal D}\underline{\alpha}\,
e^{i(\alpha_1-\alpha_2-v \alpha_3)p \cdot z} {\cal F}(\underline{\alpha})\,,
\end{align}%
where $\underline{\alpha} = \{\alpha_1,\alpha_2,\alpha_3\}$ is the set of the quark, gluon, and antiquark momentum fractions and%
\begin{align}
 \int{\cal D}\underline{\alpha} = \int_0^1 d\alpha_1 d\alpha_2 d\alpha_3 \delta(\alpha_1\!+\!\alpha_2\!+\!\alpha_3\!-\!1)\,. 
\end{align}%
C-parity implies that the DAs $\Phi$ and $\Psi$ are antisymmetric under the interchange of the quark momenta, $\alpha_1\leftrightarrow \alpha_2$, whereas $\widetilde\Phi$ and $\widetilde\Psi$ are symmetric.\par%
Taking into account contributions of the lowest and the next-to-lowest conformal spin, one obtains~\cite{Braun:1989iv}%
\begin{align}
\Phi_{4;\pi}(\underline{\alpha}) & =  120 \alpha_1\alpha_2\alpha_3 \, \phi_{1,\pi} (\alpha_1-\alpha_2) \,,
\nonumber\\
\widetilde\Phi_{4;\pi}(\underline{\alpha}) & =  120
\alpha_1\alpha_2\alpha_3 \Bigl[ \widetilde\phi_{0,\pi} + \widetilde\phi_{2,\pi} (3\alpha_3-1)\Bigr] \,,
\nonumber\displaybreak[0]\\
{\widetilde\Psi}_{4;\pi}(\underline{\alpha}) & = 
 -30 \alpha_3^2\Bigl[ \psi_{0,\pi}(1\!-\!\alpha_3)
+\psi_{1,\pi}\Bigl(\alpha_3(1\!-\!\alpha_3)-6\alpha_1\alpha_2\Bigr)
\nonumber\\&\qquad
+\psi_{2,\pi}\Bigl(\alpha_3(1\!-\!\alpha_3)-\frac{3}{2}(\alpha_1^2
                               +\alpha_2^2)\Bigr)\Bigr] \,,
\nonumber\\
 {\Psi}_{4;\pi}(\underline{\alpha}) & = 
 - 30 \alpha_3^2 (\alpha_1\!-\!\alpha_2) \Bigl[ 
                    		    \psi_{0,\pi}\! + \psi_{1,\pi} \alpha_3  
\nonumber\\&\qquad
+ \frac{1}{2} \psi_{2,\pi}(5 \alpha_3\!-\!3)\Bigr]\,.
\label{eq:T4-conformal}
\end{align}%
Omitting terms involving the twist-$3$ quark-antiquark-gluon coupling $m_u f_{3\pi}$, which are negligible, the coefficients are given by the following expressions,%
\begin{align}
  \widetilde\phi_{0,\pi} &= \psi_{0,\pi} = - \frac13\, \delta^\pi_2\,, \notag\\
  \widetilde\phi_{2,\pi}  &= \frac{21}{8} \delta^\pi_2  \omega_{4\pi} \,, \notag\\
  \phi_{1,\pi}    &=  \frac{21}{8}\left[\delta^\pi_2\omega_{4\pi}   + \frac{2}{45} m^2_\pi \left(1-\frac{18}{7}a_2^\pi\right) \right]\,,
\notag\\
 \psi_{1,\pi} & =    \frac{7}{4} \left[
  \delta^\pi_2\omega_{4\pi} \!+\! \frac{1}{45} m^2_\pi \left(1\!-\!\frac{18}{7}a_2^\pi\right) \right]\,,  
\notag\\
 \psi_{2,\pi} & =    \frac{7}{4} \left[
 2 \delta^\pi_2 \omega_{4\pi} \!-\! \frac{1}{45} m^2_\pi \left(1\!-\!\frac{18}{7}a_2^\pi\right) \right]\,,
\label{NLOspin}
\end{align}%
where $\delta^\pi_2$ and $\omega_{4\pi}$ are higher-twist parameters. The former is defined as the local matrix element%
\begin{align}
\label{def:delta-pi-2}
 \langle 0| \bar u \gamma^\rho ig \widetilde{G}_{\rho\mu} u |\pi^0(p)\rangle &= p_\mu F_\pi \delta^\pi_2\,.
\end{align}%
Its scale dependence is given by%
\begin{align}
 \delta^\pi_2(\mu)&= L^{32/(9\beta_0)} \delta^\pi_2(\mu_0)\,, 
\end{align}%
where $L = \alpha_s(\mu)/\alpha_s(\mu_0)$. It is interesting to note that, although $\omega_{4\pi}$ terms appear in individual contributions, they will cancel in the final results for all current correlations in Eqs.~\eqref{def_higher_twist}.\par%
In the diagram in Fig.~\ref{fig-pigammagamma}$(c)$, the quark propagator connecting the gluon emission point with the current gets contracted to a point, and this contribution is expressed in terms of the two-particle higher-twist DAs which are related to the three-particle DAs defined above by QCD equations of motion. The relevant techniques are explained, e.g., in Refs.~\cite{Braun:1989iv,Ball:2006wn}. One defines two-particle twist-$4$ DAs as%
\begin{align}
\MoveEqLeft[0.5]
\langle 0 | \bar u(z)[z,-z]\gamma_\mu\gamma_5 u(-z)|\pi^0(p)\rangle =
\nonumber\\
& =  i F_\pi p_\mu \int_0^1\! du \, e^{i(2u-1) p \cdot z} \Bigl[\phi_{2\pi}(u) + \frac{z^2}{4} \phi_{4\pi}(u)+\mathcal{O}(z^4)\Bigr]
\nonumber\\
&\quad +\frac{i}{2}\, F_\pi\, \frac{1}{p \cdot z}\, z_\mu\!  \int_0^1\! du \, e^{i(2u-1) p \cdot z} \bigl[\psi_{4\pi}(u)+ \mathcal{O}(z^2)\bigr]\,.
\end{align}%
Note that this equation is nothing but the light-ray OPE~\eqref{light-ray-OPE} at tree level for the two quark fields connected by the Wilson line where we retain twist-$4$ terms; the pion mass corrections arising from the application of the leading-twist projection operator to the exponential factor as in \eqref{Pi-exp} are included in $\phi_{4}(u)$. Neglecting, as above, three-particle twist-$3$ contributions $\sim m_u f_{3\pi}$ one obtains~\cite{Agaev:2014wna}%
\begin{align}
    \psi_{4\pi}(u) = \psi_{4\pi}^{{\rm twist}}(u) +  m^2_\pi \psi_{4\pi}^{{\rm mass}}(u)    
\end{align}%
with%
\begin{align}
 \psi_{4\pi}^{{\rm mass}}(u) &= \frac{17}{12} - 19 u\bar u + \frac{105}{2} u^2\bar u^2 
\nonumber\\
&\quad+  a_{2}^{\pi} \Big(\frac{3}{2} - 54  u\bar u  + 225 u^2\bar u^2 \Big)\,,
\nonumber\\
 \psi_{4\pi}^{{\rm twist}}(u) &= \frac{20}{3} \delta^\pi_2 C_2^{1/2}(2u-1) \,,
\end{align}%
and similarly%
\begin{align}
    \phi_{4\pi}(u) = \phi_{4\pi}^{{\rm twist}}(u) +  m^2_\pi \phi_{4\pi}^{{\rm mass}}(u)\,,    
\end{align}%
where%
\begin{align}
 \phi_{4\pi}^{{\rm twist}}(u) &= \frac{200}{3}\delta^\pi_2 u^2 \bar u^2  
+ 21 \delta^\pi_2 \omega_{4\pi}\Big\{ u\bar u (2\!+\!13 u\bar u)
\nonumber\\
&\quad+ 2\big[ u^3(10-15 u+6 u^2)\ln u + (u\leftrightarrow \bar u)\big]\Big\}\,,
\nonumber\\
 \phi_{4\pi}^{{\rm mass}}(u) &=  u\bar u \Big[\frac{88}{15} + \frac{39}{5} u\bar u + 14 u^2\bar u^2\Big]
\nonumber\\
&\quad-  a_2^\pi u\bar u \Big[\frac{24}{5} - \frac{54}{5} u\bar u + 180 u^2\bar u^2\Big]
+ \Big(\frac{28}{15}-\frac{24}{5} a_{2}^\pi \Big)
\nonumber\\
 &\quad\times\Big[ u^3(10-15 u+6 u^2)\ln u
 + (u\leftrightarrow \bar u)\Big] \,. \raisetag{0.5cm}
\end{align}%
Using these expressions one arrives after some algebra at the results for the higher-twist contributions to the correlations functions that are collected in the text.\par%
%
%
%\bibliography{bibliography}%
\input{DA_position_space.bbl}
\end{document}